\documentclass[prd, twocolumn, aastex]{revtex4-1}
\pdfoutput=1 
\usepackage{amsmath,amstext}
\usepackage[T1]{fontenc}
\usepackage{apjfonts} 
\usepackage{graphicx}
\usepackage{natbib}
\usepackage{xcolor}
\usepackage{hyperref}

\usepackage{aasjournal}

\begin{document}

\title{Robust and model-independent cosmological constraints from distance measurements}

\author{Zhongxu Zhai$^*$}

\author{Yun Wang}

\affiliation{IPAC, California Institute of Technology, Mail Code 314-6, 1200 E. California Blvd., Pasadena, CA 91125}

\email{zhai@ipac.caltech.edu}

\begin{abstract}
We present a systematic analysis of the cosmological constraints from the ``Pantheon Sample'' of 1048 Type Ia Supernovae (SNe Ia) in the redshift range $0.01<z<2.3$ compiled by Scolnic et al. (2018). Applying the flux-averaging method for detecting unknown systematic effects, we find that the ``Pantheon'' sample has been well calibrated and the bias caused by unknown systematic errors has been minimized. We present the estimate of distances measured from SNe Ia and reconstruct the expansion history of the Universe. The results are in agreement with a simple cosmological constant model and reveals the possible improvements that future SN Ia observations from WFIRST and LSST can target. 

We have derived distance priors using the Cosmic Microwave Background (CMB) data from the Planck 2018 final data release, and combine them with SNe Ia and baryon acoustic oscillation (BAO) data, to explore the impact from the systematic errors of SNe Ia on the combined cosmological parameter constraints. Using the combined data set of SNe Ia, BAO, and CMB distance priors, we measure the dark energy density function $X(z)=\rho_X(z)/\rho_X(0)$ as a free function (defined as a cubic spline of its values at $z=0.33, 0.67, 1.0$), along with the cosmological parameters ($\Omega_k$, $\Omega_m$, $\Omega_b$, $H_0$).
We find no deviation from a flat Universe dominated by a cosmological constant ($X(z)=1$), and $H_0=68.4\pm 0.9\text{ km s}^{-1}\text{Mpc}^{-1}$, straddling the Planck team's measurement of $H_0=67.4\pm 0.5\text{ km s}^{-1}\text{Mpc}^{-1}$, and Riess et al. (2018) measurement of $H_0=73.52\pm 1.62\text{ km s}^{-1}\text{Mpc}^{-1}$.
Adding $H_0=73.52\pm 1.62\text{ km s}^{-1}\text{Mpc}^{-1}$ as a prior to the combined data set leads to the time dependence of the dark energy density at $z\sim 0.33$ at 68\% confidence level. Not including the systematic errors on SNe Ia has a similar but larger effect on the dark energy density measurement.
\end{abstract}


\maketitle

\section{Introduction}

The accelerated expansion of the Universe has been one of the greatest mysteries in modern cosmology since it was first discovered through the observations of supernovae \citep{Riess_1998, Perlmutter_1999}. The interpretation of this phenomena motivates the concept of dark energy, which dominates the total energy in the present-day observable Universe. As the mathematically simplest solution, the cosmological constant can well match the current observations. But there are arguments in the literature that the state-of-art observations imply deviations from the cosmological constant, e.g. \cite{Zhao_2017, Sola_2017a, Park_2018a, Park_2018b, Wangyuting_2018} and references therein. Therefore it is necessary to perform a systematic and consistent analysis of the observational data so as to arrive at robust and model-independent constraints on the properties of dark energy \citep{Wang_2016}. 

In this paper, we investigate the possible unknown systematic uncertainties in the latest observational data from Type Ia supernovae (SNe Ia), the $``$Pantheon Sample$"$ which consists of 1048 SNe Ia in the redshift range $0.01<z<2.3$ \citep{Scolnic_2018}. On the other hand, the Planck team has released the final product of the cosmic microwave background (CMB) measurement from the Planck mission \citep{Planck_2018_1, Planck_2018_6}, which improves the constraints from the previous releases and favors a spatially-flat 6-parameter $\Lambda$CDM cosmology. Additionally, galaxy clustering measures the distance scales through the Baryon Acoustic Oscillation (BAO) at various redshifts \citep{Eisenstein_2005}. These measurements probe the Universe with different methodologies, therefore they have different systematic uncertainties, and a consistency check is necessary to ensure a robust constraint on dark energy \cite{Wang_2004}. On the other hand, the CMB observations from the Planck mission indicate a measurement on the Hubble constant $H_{0}$ that is in tension with the local measurement from the distance ladder. Compared with the latest measurement using data from Gaia \citep{Riess_2018b}, this discrepancy is higher than 3.6$\sigma$. Whether this discrepancy implies the unknown systematic error in either observations, or signifies novel physics beyond our current understanding of cosmology is important for the survey strategies in the next decades. In this paper, we also investigate the impact of imposing this latest $H_{0}$ distance ladder measurement as a prior on the dark energy constraint.

Our paper is organized as follows. In Section 2, we outline the models of dark energy we use in the analysis. Section 3 presents our analysis of the Pantheon SNe Ia data, including the search for unknown systematic errors using the flux-averaging method, and the measurement of the cosmic expansion history. In Section 4, we show results from combinations of different observational datasets. We discuss and conclude in Section 5.

\section{Methodology}

We focus on the the distance measurements of the Universe in this paper, i.e. the geometrical expansion history of the Universe, thus we do not consider the contribution from the growth rate of the cosmic large scale structure. The model is based on a FRW metric, under which the comoving distance to an object at redshift $z$ is given by
\begin{eqnarray}\label{eq:comodis}
    r(z) = \frac{c}{H_{0}}|\Omega_{k}|^{-1/2}\text{sinn}[|\Omega_{k}|^{1/2}\Gamma(z)], \\
    \Gamma(z) = \int_{0}^{z}\frac{dz'}{E(z')}, \quad E(z) = H(z)/H_{0},
\end{eqnarray}
where $c$ is the speed of light, $H_{0}=100 h\text{ km s}^{-1}\text{Mpc}^{-1}$ with $h$ the dimensionless Hubble constant, $\text{sinn}(x)=\sin(x), x, \text{and } \sinh(x)$ for $\Omega_{k}<0, \Omega_{k}=0, \text{and } \Omega_{k}>0$, respectively. The Hubble parameter $H(z)$ is given by
\begin{equation}
    H(z)^2 = H_{0}^{2}[\Omega_{m}(1+z)^3+\Omega_{r}(1+z)^4 + \Omega_{k}(1+z)^2 + \Omega_{\Lambda}X(z)],
\end{equation}
with constraint $\Omega_{m}+\Omega_{r}+\Omega_{k}+\Omega_{\Lambda}=1$. The function $X(z)\equiv\rho_{X}(z)/\rho_{X}(0)$ describes the evolution of dark energy density with time. The radiation term $\Omega_{r}=\Omega_{m}/(1+z_{\text{eq}})\ll\Omega_{m}$ can be omitted in the late time studies of dark energy and $z_{\text{eq}}$ is the redshift of matter-radiation equality. $\Omega_{k}$ denotes the contribution from spatial curvature. 

The cosmological constant corresponds to $\rho_{X}=const$. For comparison with this simplistic model, we also consider parameterizations of the equation of state of dark energy $w(z)$
\begin{equation}
    \frac{\rho_{X}(z)}{\rho_{X}(0)}=(1+z)^{3(1+w(z))}.
\end{equation}
The first model is to assume $w(z)=const$ (hereafter $w$CDM model). In this case, the $\Lambda$CDM model is a special class when $w=-1$. The second model is linear parameterization of the cosmic scale factor $a=1/(1+z)$ \cite{Chevallier_2001, Linder_2003} (hereafter $w_{0}w_{a}$CDM model)
\begin{equation}
    w(z)=w_{0} + (1-a)w_{a},
\end{equation}
where $w_{0}$ and $w_{a}$ are free parameters to be determined by observational data. Note that $w$CDM and $\Lambda$CDM model are just special classes when $w_{0}$ and $w_{a}$ are fixed with particular values.

Finally, we consider a model-independent parameterization of $X(z)=\rho_{X}(z)/\rho_{X}(0)$, where $X(z)$ is a free function of redshift given by the cubic spline of its values at $z=1/3, 2/3$ and $1.0$ and assume that $X(z>1)=X(z=1)$. This model has been investigated by \cite{Wang_2009, Wang_2016}.

\section{Analysis of Type Ia Supernovae data}

\subsection{Flux averaging}\label{sec:FA}

The published SNe data are usually analyzed in terms of the distance modulus
\begin{equation}
    \mu_{0} \equiv m-M = 5\log{\left[\frac{d_{L(z)}}{\text{Mpc}}\right]}+25,
\end{equation}
where $m$ and $M$ are apparent and absolute magnitude of each supernova respectively, and the luminosity distance is given by $d_{L}(z)=(1+z)r(z)$. 

The SNe dateset considered in this paper is the Pantheon sample. This compilation contains the full set of specctroscopically confirmed Pan-STARRS1 (PS1) SNe Ia and in combination with spectroscopically confirmed SNe Ia from CfA1-4, CSP, PS1, SDSS, SNLS and \emph{Hubble Space Telescope (HST)} SN surveys \citep{Scolnic_2018}. The (cross-)calibration of these SN samples can reduce the systematics substantially and the details be found in \cite{Betoule_2013, Scolnic_2015, Scolnic_2018} and references therein. 

Due to the degeneracy between $H_{0}$ and the absolute magnitude, the Pantheon sample reports the corrected magnitude for each SN. In the cosmological analysis, this can be marginalized by the method presented in \cite{Conley_2011}. In addition, we use the unbinned, full SN data set instead of the binned version to be in line with general community reproducibility \citep{Scolnic_2018}

In the cosmological analysis of SNe data, one of the main systematics is the weak gravitational lensing induced by galaxies because of the inhomogeneous distribution of matter in the Universe \citep{Wang_2004}. One way to remove or reduce gravitational-lensing bias is through flux averaging \citep{Wang_2004}. Because of flux conservation, the average magnification of a sufficient number of SNe at the same redshift is unity. And thus this process can recover the unlensed brightness of the SNe and yield cosmological parameter estimations without bias from weak gravitational lensing \citep{Wang_2000}. An important additional benefit of flux-averaging is that it effectively reduces a global systematic bias (over the entire redshift range) into a local bias (within each redshift bin) with a much smaller amplitude. Thus flux-averaged SN Ia data should be significantly less affected by unknown systematic biases (such as weak lensing or imperfect K-correction) than SNe Ia data without flux-averaging \cite{Wang_2005}.

We explore the cosmological parameter constraints through the $\chi^2$ statistics combined with Markov Chain Monte Carlo (MCMC). Our flux-averaging method follows \cite{Wang_2016} and the steps are summarized below:

1. Convert the distance modulus of SNe Ia into ``fluxes$"$:
\begin{equation}
    F(z_{l}) = 10^{(\mu_{0}^{\text{data}}(z_{l})-25)/2.5}=\Big(\frac{d_{L}^{\text{data}}(z_{l})}{\text{Mpc}}\Big)^{-2}.
\end{equation}

2. Remove the redshift dependence of these ``fluxes$"$ to obtain their ``absolute luminosities$"$  ${\mathcal{L}(z_{l})}$ by assuming a set of cosmological parameters ${\bf{s}}$,
\begin{equation}
    \mathcal{L}(z_{l)} \equiv F(z_{l}) d_{L}^2(z_{l}|\bf{s}).
\end{equation}

3. Flux-average the ``absolute luminosities$"$ ${\mathcal{L}(z_{l})}$ in each redshift bin $i$ to obtain $\bar{\mathcal{L}^{i}}$ and mean redshift
\begin{equation}
    \bar{\mathcal{L}^{i}}=\frac{1}{N_{i}}\sum_{i=1}^{N_{i}}\mathcal{L}_{l}^{i}(z_{l}^{(i)}), \qquad \bar{z_{i}} = \frac{1}{N_{i}}\sum_{l=1}^{N_{i}}z_{l}^{(i)},
\end{equation}
where $N_{i}$ is the number of SNe in $i-$th redshift bin.

4. Place $\bar{\mathcal{L}^{i}}$ at the mean redshift $\bar{z_{i}}$ to get the binned flux
\begin{equation}
    \bar{F}(\bar{z_{i}}) = \bar{\mathcal{L}^{i}}/d_{L}^2(\bar{z_{i}}|\bf{s}),
\end{equation}
and the flux-averaged distance modulus
\begin{equation}
    \bar{\mu}^{\text{data}}(\bar{z_{i}}) = -2.5\log_{10}\bar{F}(\bar{z_{i}}) + 25.
\end{equation}

5. The new covariance matrix of $\bar{\mu}(\bar{z_{i}})$ and $\bar{\mu}(\bar{z_{j}})$ can be computed by
\begin{eqnarray}
       && \text{Cov}[\bar{\mu}(\bar{z_{i}}), \bar{\mu}(\bar{z_{j}})] = \frac{1}{N_{i}N_{j}\bar{\mathcal{L}^{i}}\bar{\mathcal{L}^{j}}}\cdot \\
       &&\sum_{l=1}^{N_{i}}\sum_{m=1}^{N_{j}}\mathcal{L}(z_{l}^{(i)})\mathcal{L}(z_{m}^{(j)})\langle \Delta\mu_{0}^{\text{data}}(z_{l}^{(i)}) \Delta\mu_{0}^{\text{data}}(z_{m}^{(j)}) \rangle,
\end{eqnarray}
where $\langle \Delta\mu_{0}^{\text{data}}(z_{l}^{(i)}) \Delta\mu_{0}^{\text{data}}(z_{m}^{(j)}) \rangle$ is the covariance between the corresponding SN Ia pairs from the measured distance moduli.

6. The final $\chi^2$ of this flux-averaged data ${\bar{\mu}}(\bar{z}_{i})$ can be computed as
\begin{equation}
    \chi^2 = \sum_{ij}\Delta\bar{\mu}(\bar{z}_{i})\text{Cov}^{-1}[\bar{\mu}(\bar{z_{i}}), \bar{\mu}(\bar{z_{j}})] \Delta\bar{\mu}(\bar{z}_{j}),
\end{equation}
where $\Delta\bar{\mu}(\bar{z}_{i})\equiv \bar{\mu}(\bar{z}_{i})-\mu^{p}(\bar{z}_{i}|\bf{s})$, and $\mu^{p}(\bar{z}_{i}|\bf{s})$ can be computed for a given cosmological model.

\begin{figure*}[htbp]
\begin{center}
\includegraphics[width=8.5cm]{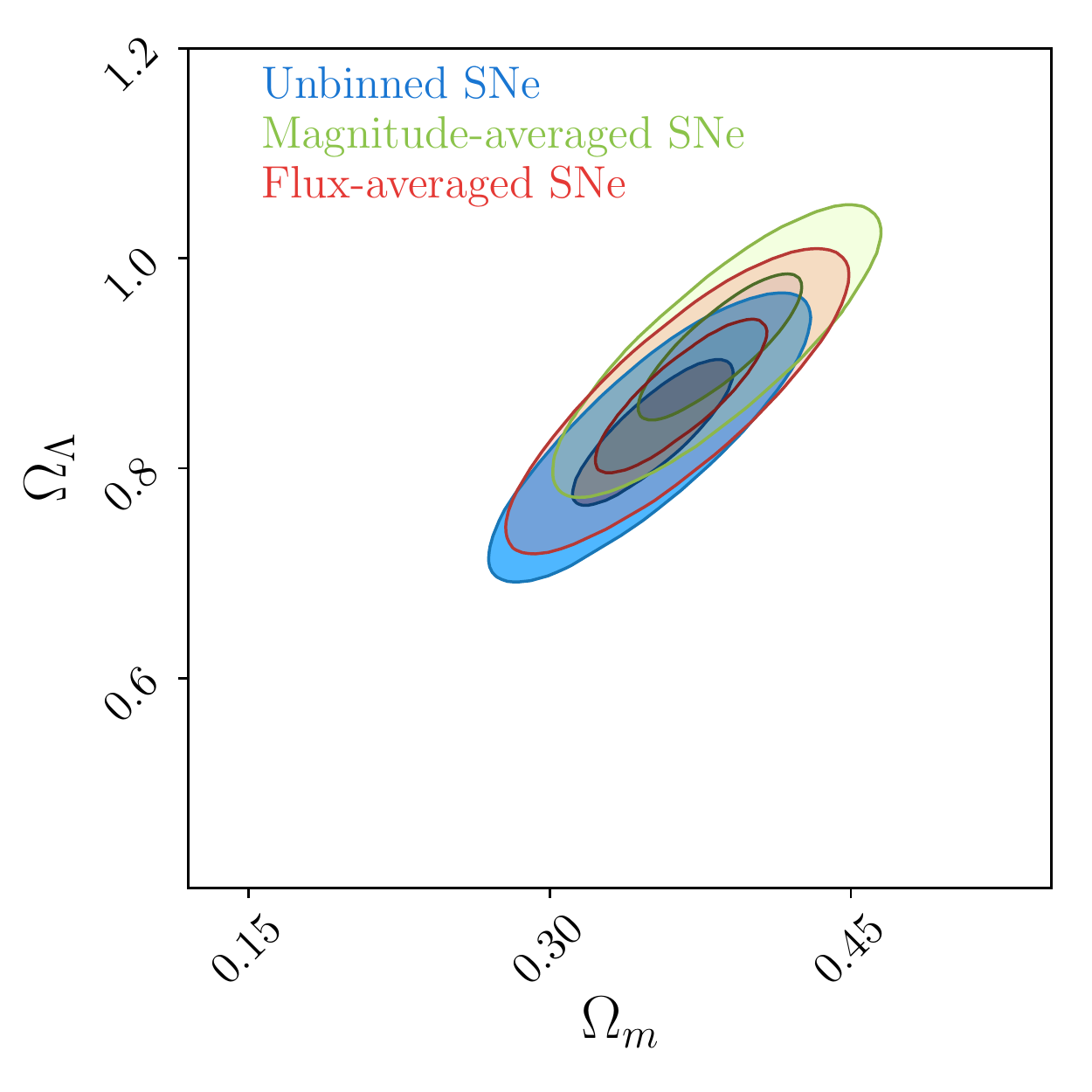}
\includegraphics[width=8.5cm]{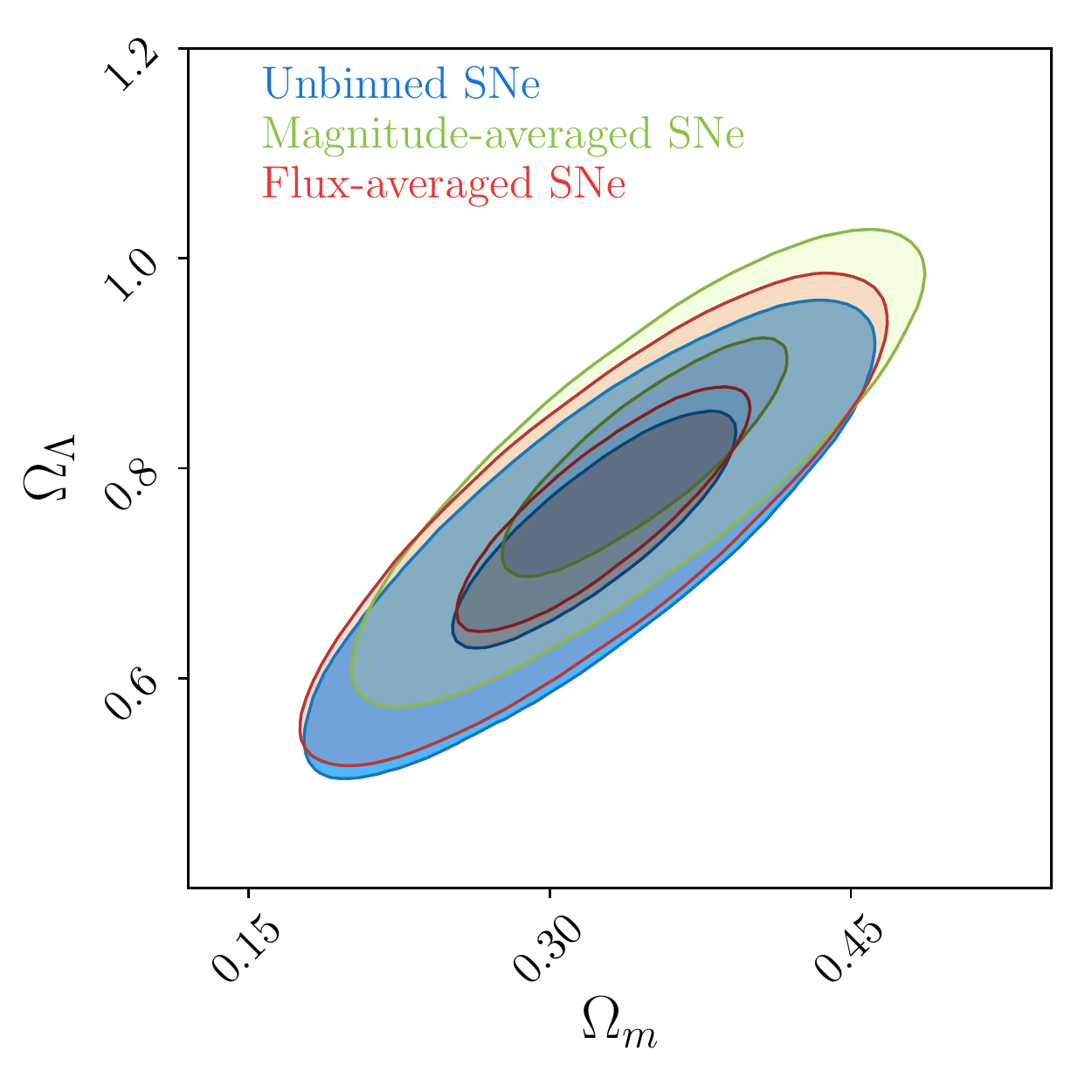} \\
\includegraphics[width=8.5cm]{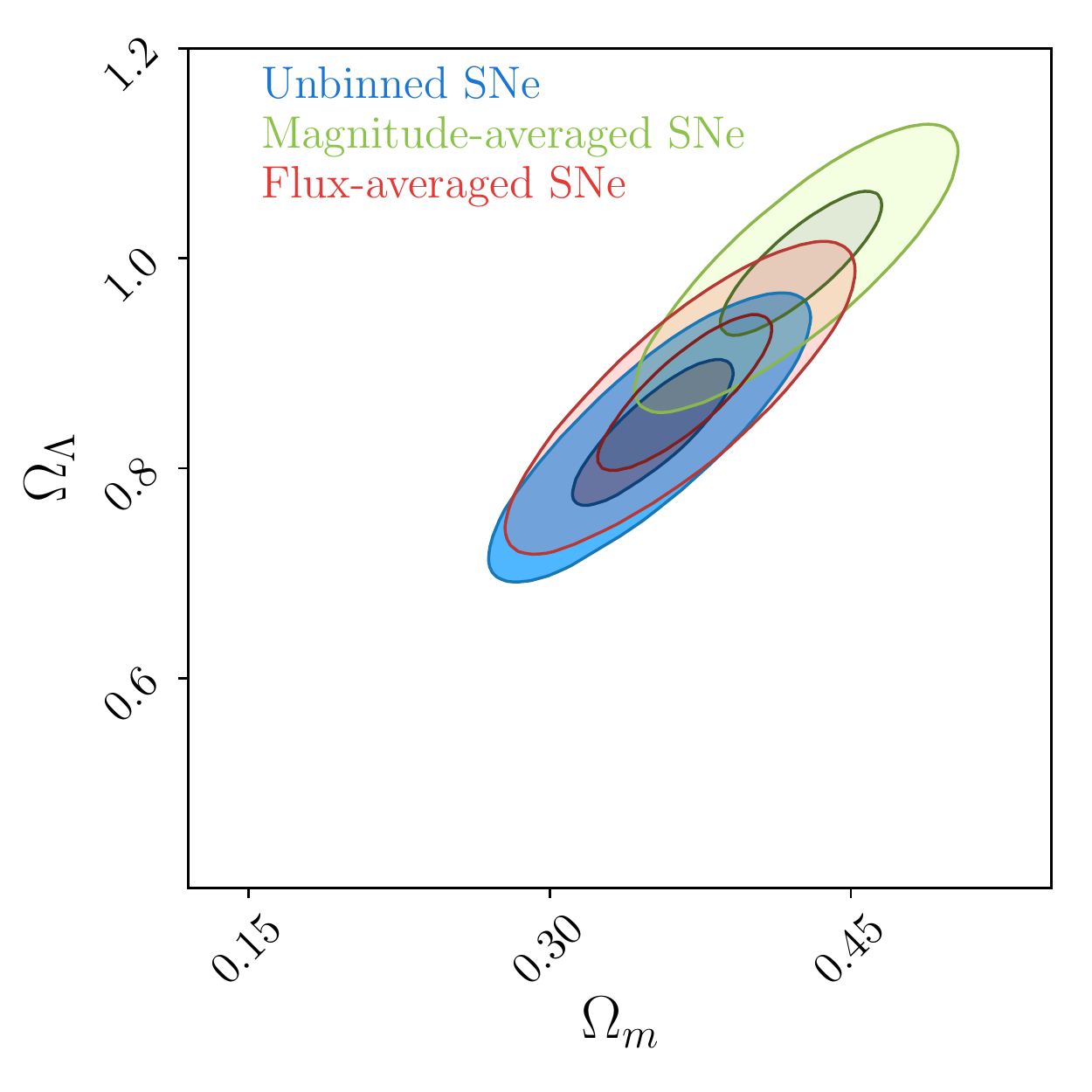}
\includegraphics[width=8.5cm]{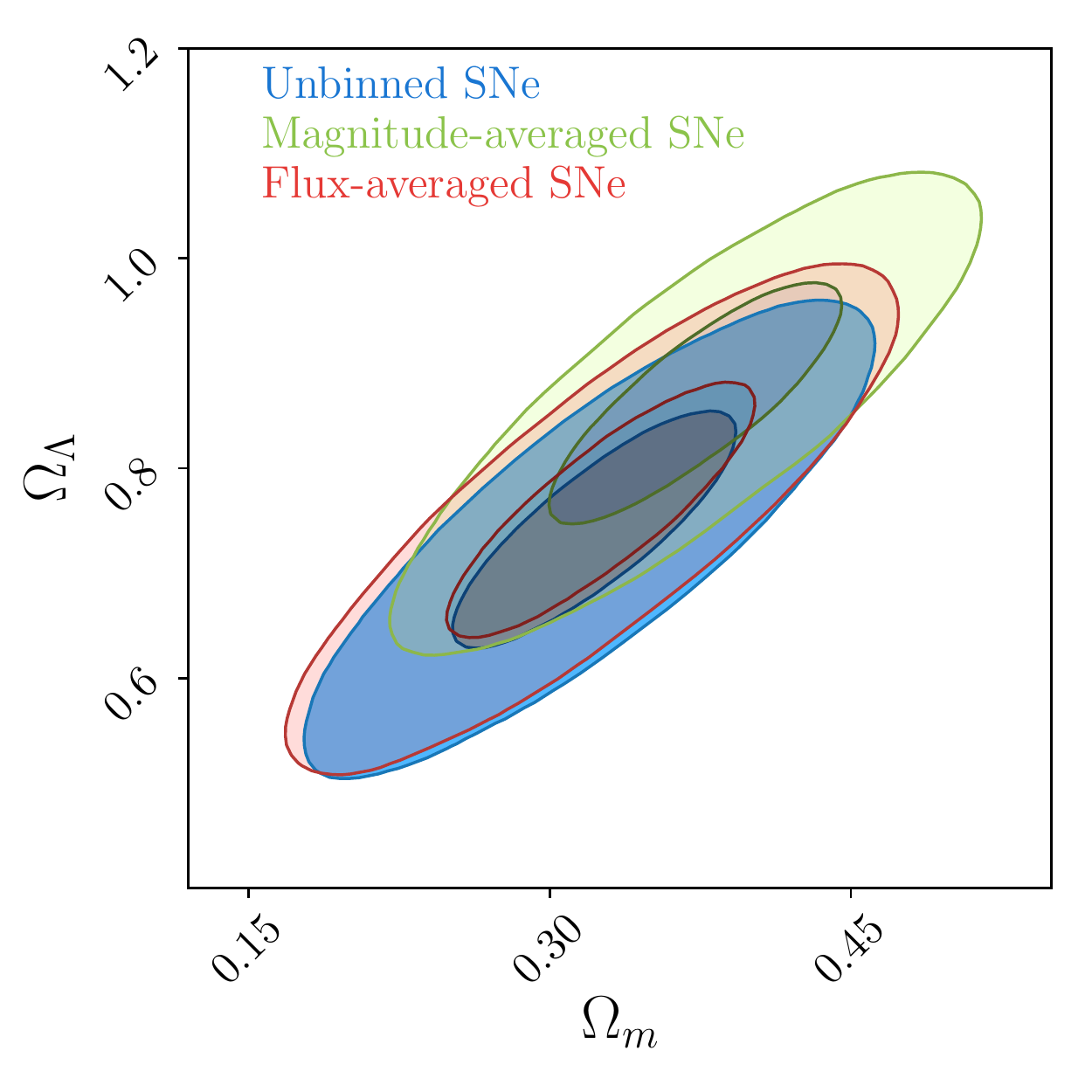}
\caption{Parameter constraints for a non-flat $\Lambda$CDM cosmology with full unbinned Pantheon sample (blue), flux-averaged sample (red) and magnitude-averaged sample (green). The left and right panels assume analysis without and with systematic uncertainties respectively. The top row adopts the same redshift binning scheme as the Pantheon paper provides which has 40 bins, the bottom row adopt the binning scheme with a downsampling factor of 2, resulting 20 bins.}
\label{fig:FA_oLCDM}
\end{center}
\end{figure*}

\begin{figure*}[htbp]
\begin{center}
\includegraphics[width=8.5cm]{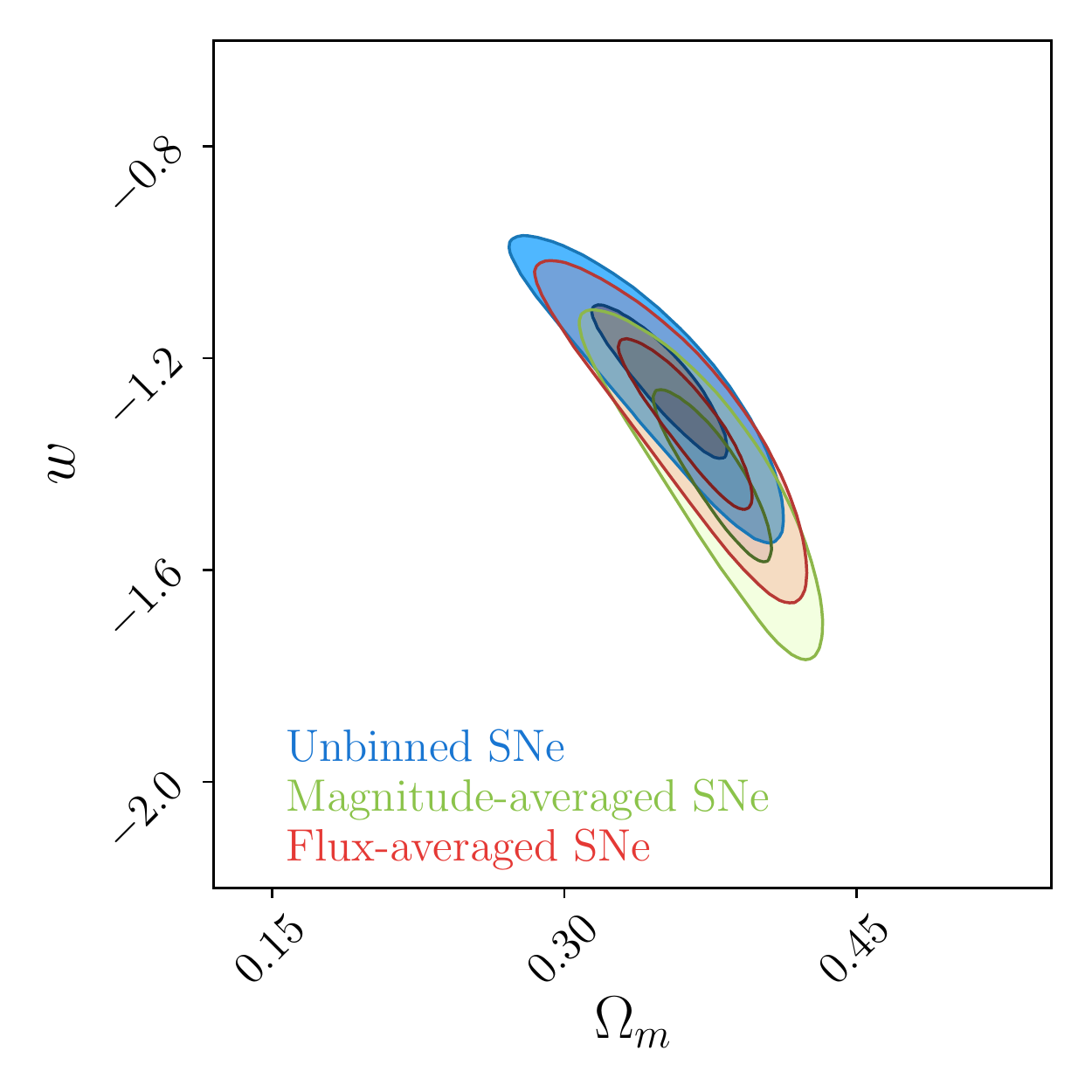}
\includegraphics[width=8.5cm]{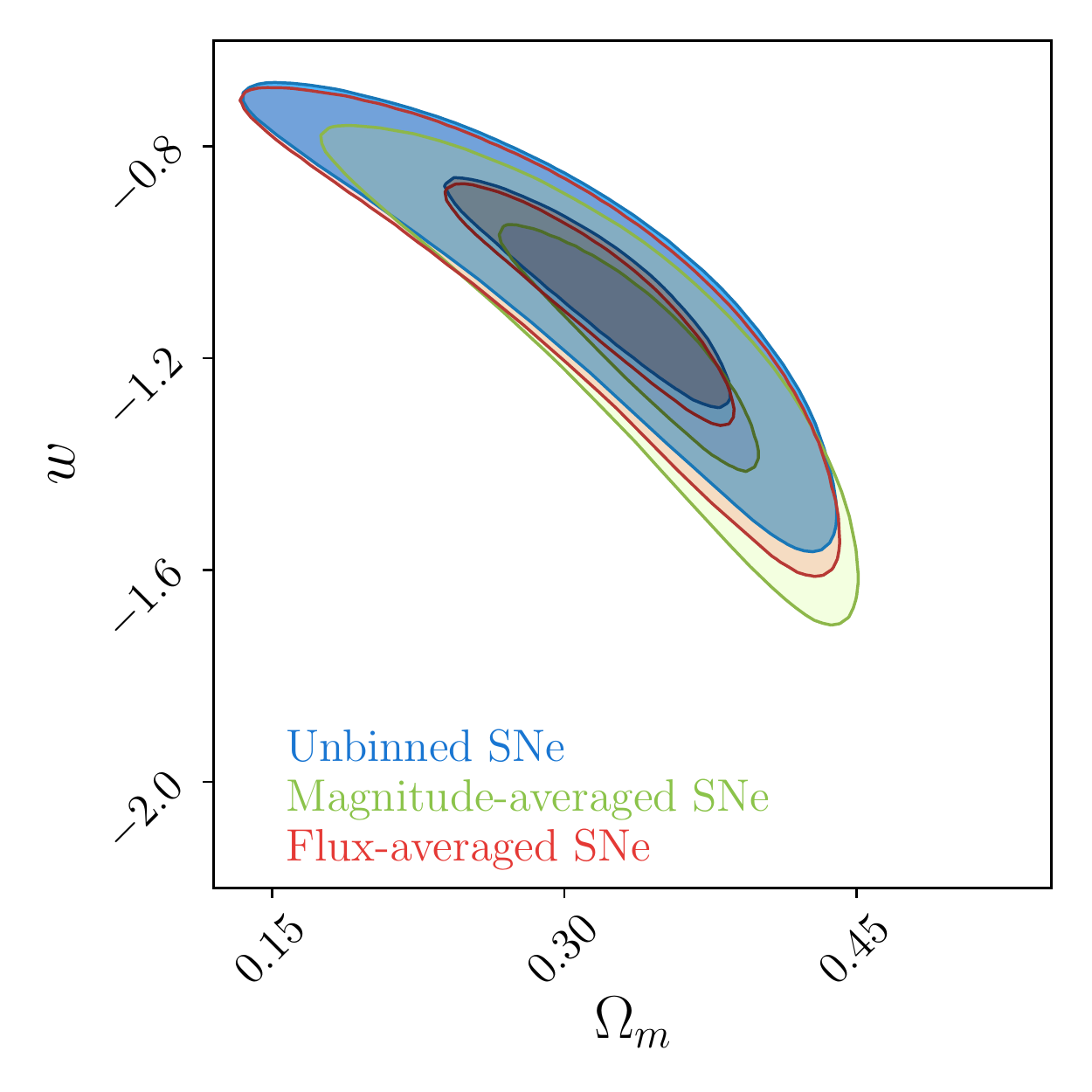} \\
\includegraphics[width=8.5cm]{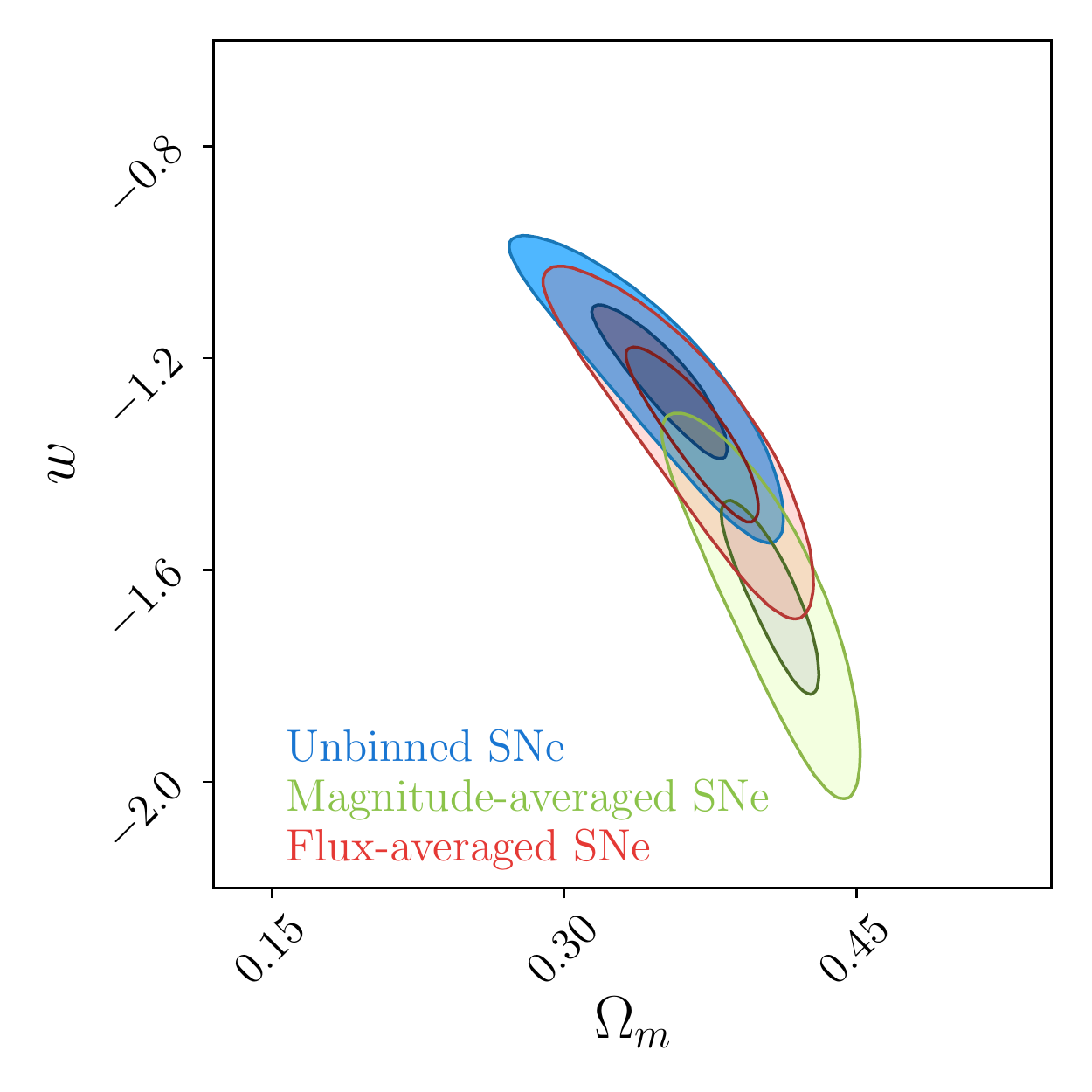}
\includegraphics[width=8.5cm]{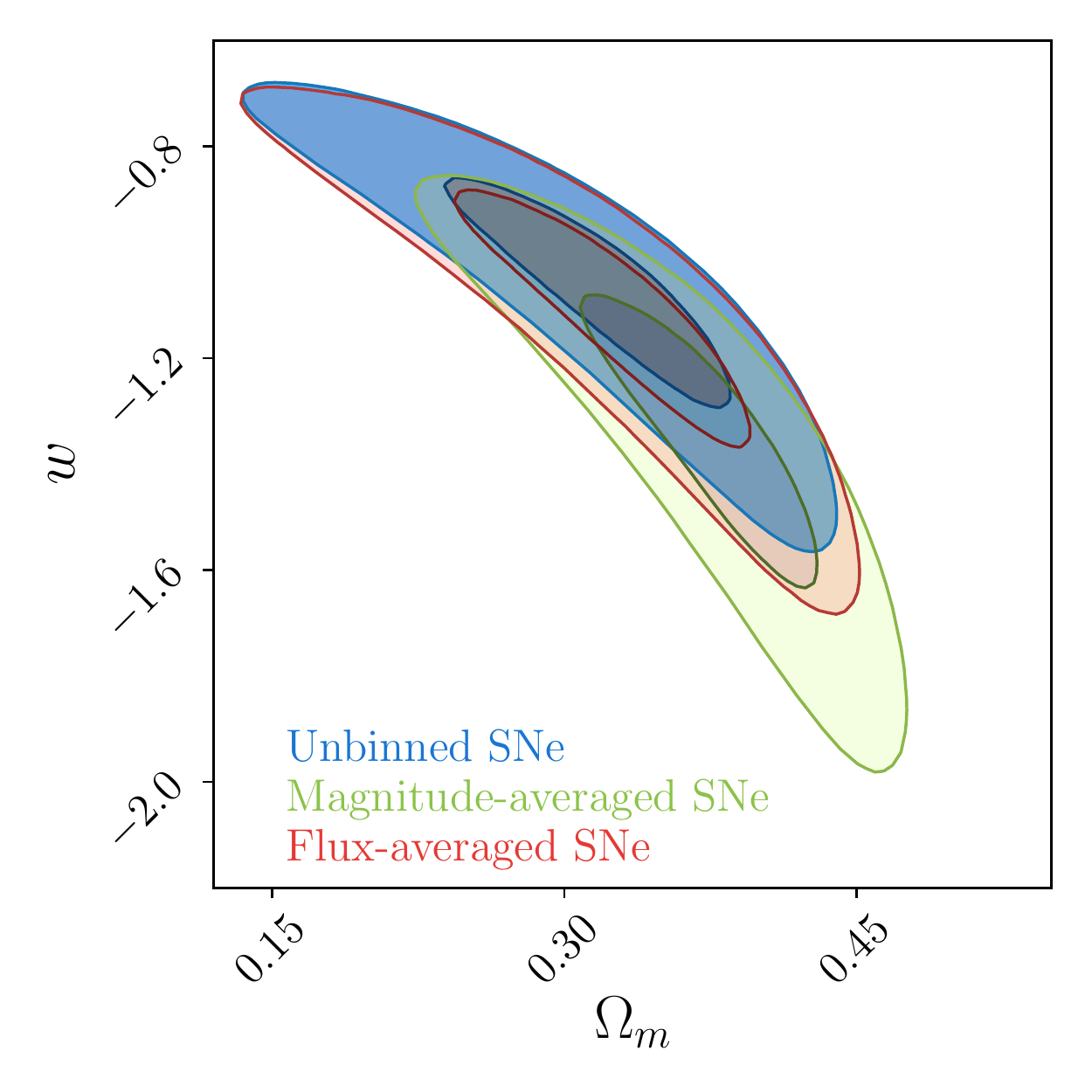}
\caption{The same as Figure \ref{fig:FA_oLCDM} but for flat $w$CDM cosmology.}
\label{fig:FA_XCDM}
\end{center}
\end{figure*}

In order to compare the effect of flux-averaging of SNe, we also perform a ``magnitude-averaging$"$ analysis. In this framework, the ``magnitude-averaged$"$ distance moduli in the $i-$th redshift bin is just the mean of all the SNe in this redshift bin. The results of flux-averaging and magnitude-averaging are compared with the unbinned full Pantheon sample, and any systematic bias can be revealed by the offset in the cosmological parameter constraints. In our calculation, we adopt the same redshift binning scheme as the binned Pantheon sample and explore effect of binning width. We apply these methods to non-flat $\Lambda$CDM model and flat $w$CDM model, which has parameter set $(\Omega_{m}, \Omega_{\Lambda})$ and $(\Omega_{m}, w)$ respectively. Our results are shown in Figure \ref{fig:FA_oLCDM} and Figure \ref{fig:FA_XCDM}. In order to explore the impact from systematic error of supernovae data, we also show results with and without adding in the systematic covariance matrix in the analysis. The result is consistent with the analysis in \cite{Scolnic_2018}: the systematic uncertainties degrade the constraints on the cosmological parameters and also induce a shift in the best-fit values. This is true for both models in the figures and the shift can be attributed to the systematics in the low redshift data \citep{Scolnic_2018}. A detailed discussion of the effects and origins of systematic uncertainties can be found in the Pantheon paper \citep{Scolnic_2018} as well.

The SNe sample with flux-averaging shows very consistent constraints with the unbinned full sample as we can see from the figures and this is true for both models. The best-fit values have a offset much smaller than $1\sigma$ especially when the systematic uncertainties are taken into account. This result also shows that the original Pantheon sample has been very well cleaned and the bias induced by unknown systematic effects has been minimized. Since the weak lensing effect cannot be removed, this indicates that weak lensing effects are small for the Pantheon sample.
On the other hand, the SNe sample with magnitude-averaging shows obvious offset in the parameter constraints. It is more significant when we increase the width of the redshift bin which results a $>1\sigma$ tension for both $\Lambda$CDM and $w$CDM models when only the statistical error is applied. The addition of systematic uncertainties can alleviate the tension for magnitude-averaging at some extent but the difference is still quite noticeable. The results of flux-averaging and magnitude-averaging are quite different because different quantities are averaged in a given redshift bin. The quantity averaged in flux-averaging is an ``absolute luminosity$"$ which doesn't have redshift dependence and is thus independent from redshift binning. However, this is not true for magnitude-averaging and may induce redshift-binning dependent bias in the cosmological constraints. The consistency between the flux-averaing method and the full unbinned result shows that flux-averaging can give robust cosmological constraints, and the flux-averaged SN data can also serve as an alternative to compress the full SNe data for cosmological analysis \citep{Betoule_2014}.

\subsection{Model-independent distance measurements from SNe Ia}

\begin{figure}[htbp]
\begin{center}
\includegraphics[width=9.0cm]{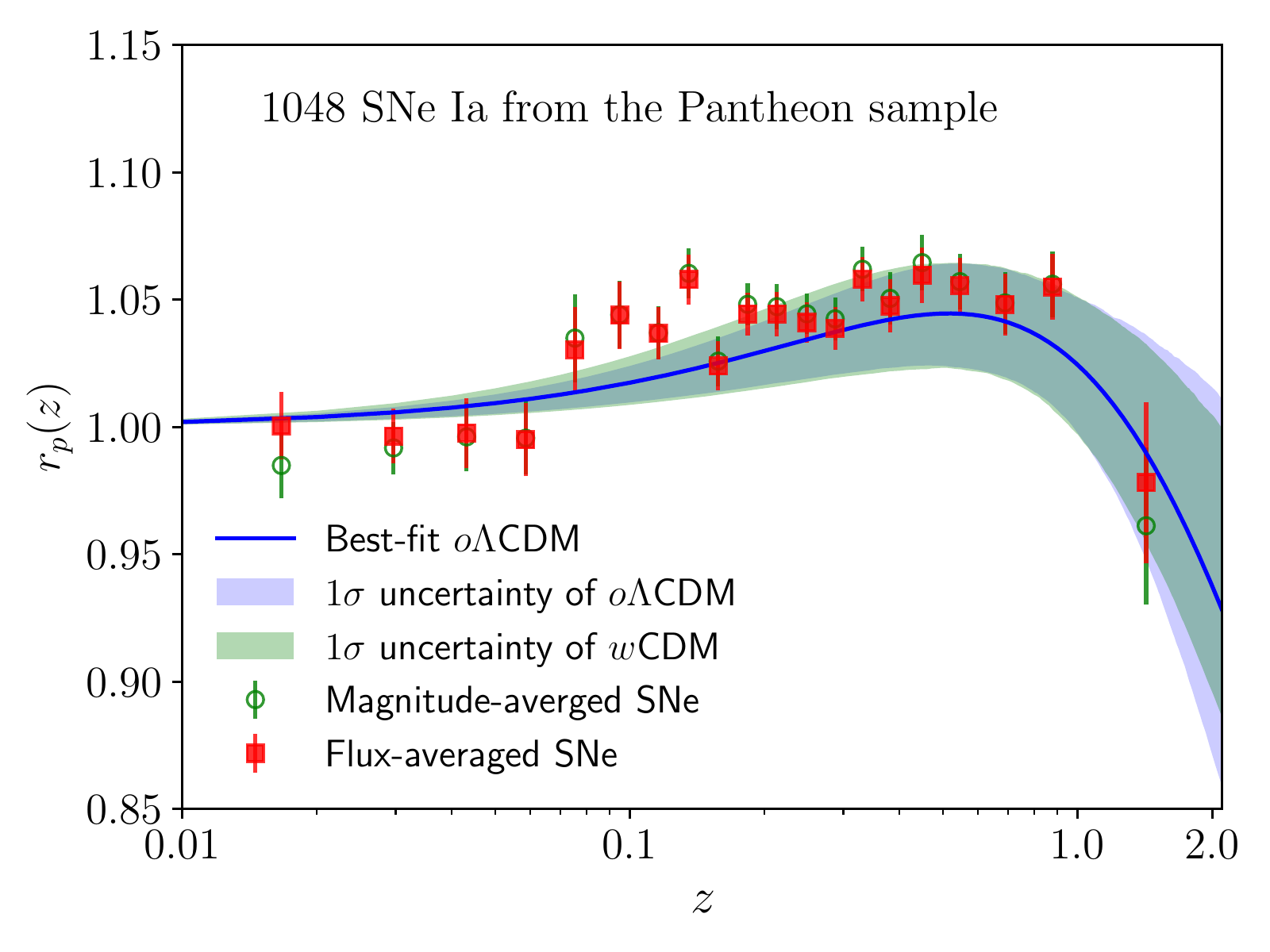}
\caption{Distance measurements from the SNe Ia dataset. The model prediction from the best-fit $o\Lambda$CDM model is shown as the blue solid line. The shaded areas represent the $1\sigma$ uncertainty from the MCMC chains.}
\label{fig:rp}
\end{center}
\end{figure}

In order to obtain an intuitive understanding of the SNe Ia data, \cite{Wang_2009} propose a scaled distance measure
\begin{equation}\label{eq:rp}
    r_{p}(z) = \frac{r_{z}}{cH_{0}^{-1}z}(1+z)^{0.41},
\end{equation}
where $r(z)$ is the comoving distance given by Eq. (\ref{eq:comodis}). This scaled distance can be measured from the SNe data, and the value at arbitrary redshift is given by cubic-spline interpolation, without explicitly assuming a cosmological model. The power 0.41 is chosen to make this distance as flat as possible over the redshift range of interest \citep{Wang_2009}. This scaled distance also provides a way to examine the difference of flux-averaging and magnitude-averaging methods. It is worth pointing out that the Pantheon sample reports corrected distance modulus instead of a raw measurements. It won't affect the cosmological constraints due to the marginalization algorithm. However it should be accounted for in the distance measurements. In our calculation, we assume it can be corrected by a constant offset between the reported distance modulus and the model prediction. We find this offset by a simple linear fit with a best-fit cosmology constrained by the full unbinned dataset. This model is also used to obtain the flux-averaged and magnitude-averaged SNe data. With 40 redshift bins, our measurements of $r_{p}$ are shown in Figure \ref{fig:rp}. For comparison, the theoretical predictions from the constrained non-flat $\Lambda$CDM and flat $w$CDM are also shown.

Compared with the model prediction, the flux-averaged distance measurements have smaller dispersion than the magnitude-averaged data. It is consistent with the fact that flux-averaging can return better $\chi^2$ in the data-model fitting \cite{Wang_2004}. In addition, the figure shows that the data points have deviations from the $\Lambda$CDM and $w$CDM cosmology at redshift $z\sim0.1$, implying that future observations at this redshift range will be very informative and provide useful cosmological constraints.

\subsection{Cosmic expansion history from SNe Ia}

The SNe data can constrain the cosmic expansion history $H(z)$ in principle. However the optimal way of doing this is still challenging in practice \citep{Tegmark_2002}. With the distance modulus measured from SNe, \cite{Wang_2005} present a optimal method to measure $H(z)$ in uncorrelated redshift bins. Since the latest Pantheon sample is much larger than the datasets analyzed in those works, it is worth revisiting this problem and investigate the implications. Note that the method in \cite{Wang_2005} assumes that the SNe measurements are uncorrelated. This condition is satisfied when we only consider the statistical uncertainty. Taking into account the systematic uncertainties may degrade the robustness of this analysis due to the correlated nature of the systematic uncertainties. In our calculation, we ignore the off-diagonal elements in the systematic covariance matrix and it may not have significant effect since the systematic uncertainty is in general much smaller than the statistical uncertainty in the Pantheon sample.

Our method to measure the cosmic expansion history is described in detail by \cite{Wang_2005} and we summarize below. We first convert the distance modulus $\mu_{0}$ of the SNe to comoving distance through
\begin{equation}
    \frac{r(z)}{1\text{Mpc}}=\frac{1}{2997.9(1+z)}10^{\mu_{0}/5-5}.
\end{equation}
And for individual SN, its comoving distance is added by a noise term
\begin{equation}
    r_{i} = r(z_{i})+n_{i},
\end{equation}
where the noise $n_{i}$ is drawn from a Gaussian distribution with zero mean, and dispersion $\sigma_{i}$ calculated from the error propagation of $\mu_{0,i}$.

The next step is to sort the SNe by increasing redshift and define the quantities
\begin{eqnarray}
    x_{i}&\equiv&\frac{r_{i+1}-r_{i}}{z_{i+1}-z_{i}}=\frac{\int_{z_{i}}^{z_{i+1}}\frac{dz'}{H(z')}+n_{i+1}-n_{i}}{z_{i+1}-z_{i}}\\
    &=&\bar{f_{i}}+\frac{n_{i+1}-n_{i}}{\Delta z_{i}},
\end{eqnarray}
where $\Delta z_{i}=z_{i+1}-z_{i}$ and $\bar{f_{i}}$ is the average of $1/H(z)$ over the redshift range $(z_{i}, z_{i+1})$. We note that in the Pantheon sample, there are several SN Ia pairs that have the same redshift, which prevents the redshift sorting. As a conservative choice, we just select the SN with larger uncertainty. In the above model, $x_{i}$ gives an unbiased estimate of the average of $1/H(z)$ in the redshift bin, since $\langle x_{i}\rangle=\bar{f_{i}}$. Thus the inverse can be used to probe the cosmic expansion history. The covariance matrix of $x_{i}$ can be calculated as
\begin{eqnarray}
    & N_{i, i-1}=-\frac{\sigma_{i}^2}{\Delta z_{i-1}\Delta z_{i}}, \qquad N_{i,i}=\frac{\sigma_{i}^2+\sigma_{i+1}^2}{\Delta z_{i}^2},\\ 
    & N_{i,i+1}=-\frac{\sigma_{i+1}^2}{\Delta z_{i}\Delta z_{i+1}},
\end{eqnarray}
with all other entries to be zero. This new data vector $x_{i}$ and the covariance matrix express the cosmic expansion history in very fine redshift bins, but with significant noise contribution. Therefore the next step is to average these noisy measurements into minimum-variance measurements in wider redshift bins. We use ${\bf{x}}^{b}$ to denote the subset of $x_{i}$ in the $b$-th redshift bin and ${\bf{N}}_{b}$ the corresponding covariance matrix, then their weighted average $y_{b}$ can be written as
\begin{equation}
    y_{b}={\bf{w}}^{b}\cdot{\bf{x}}^{b}
\end{equation}
with some weight vector ${\bf{w}}^{b}$. This weight vector can be obtained by minimizing the variance
\begin{equation}\label{eq:yb_var}
    \Delta y_{b}^2\equiv\langle y_{b}^2\rangle - \langle y_{b}\rangle^2 = {\bf{w}}^{bt}{\bf{N}}_{b}{\bf{w}}^{b},
\end{equation}
subject to the constraint that the weights add up to unity, where the superscript $t$ refers to matrix transpose. The solution to this problem by the use of the Lagrange multiplier method is 
\begin{equation}\label{eq:window}
    {\bf{w}}^{b}=\frac{{\text{\bf{N}}}^{-1}{\text{\bf{e}}}}{{\text{\bf{e}}}^{t}{\text{\bf{N}}}^{-1}{\text{\bf{e}}}},
\end{equation}
where ${\text{\bf{e}}}$ is the unit vector with all elements equal to 1. This weight vector is also called $window$ $function$, since it shows the contribution to the measurements $y_{b}$ from different redshifts. The uncertainty of measurement $y_{b}$ can be calculated by substituting the weight vector to Eq (\ref{eq:yb_var})
\begin{equation}
    \Delta y_{b} = ({\text{\bf{e}}}^{t}{\text{\bf{N}}}^{-1}{\text{\bf{e}}})^{-1/2}.
\end{equation}

\begin{figure}[htbp]
\begin{center}
\includegraphics[width=8.5cm]{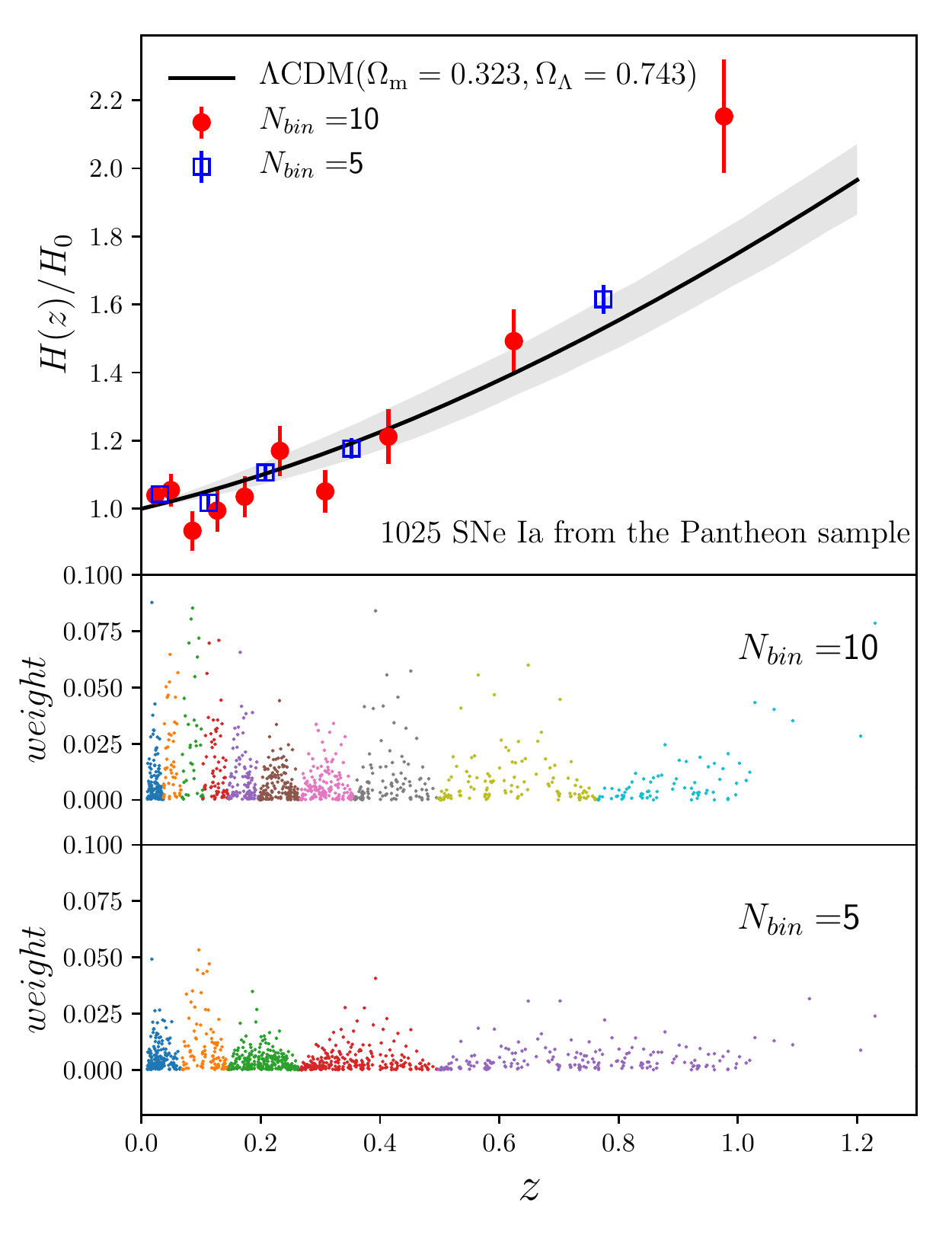}
\caption{The top panel shows the cosmic expansion history expressed in terms of dimensionless Hubble parameter, using the reconstruction method for SNe data. The solid line and grey area is the best-fit $\Lambda$CDM model and 95 $\%$ uncertainty. The measurements from data are shown with two redshift binning schemes which have 10 and 5 redshift bins respectively. The middle and bottom panels present the corresponding window functions in the reconstruction.}
\label{fig:Ez}
\end{center}
\end{figure}

Figure \ref{fig:Ez} shows our measurements of the expansion history through this method for two different redshift binning schemes, as well as the prediction from the best-fit model and 95 $\%$ uncertainties. While the reconstruction result is in agreement with the $\Lambda$CDM model, the high redshift bin has apparent deviation. This is due to the limited number of SNe in this redshift bin, and can be improved when multiple bins are combined as shown by the blue square at high redshift. The middle and bottom panels show the window functions Eq (\ref{eq:window}) of the reconstructions. Due to noisy nature of the SNe dataset, the window function only shows weak characteristics as presented in \cite{Wang_2005}: an upside-down parabola vanishing at the bin end points and maximize near the center of the bin.
Clearly, the measurement of $H(z)$ from SNe Ia can be significantly improved with the $z<0.8$ SNe Ia from LSST\cite{LSST-sciece-book}, and the $z>1$ SNe Ia from WFIRST \cite{Spergel_2015}.

\section{Combination with other datasets}

In order to improve the cosmological constraints, it is useful to include measurements from other probes. In this paper, we employ the data from CMB, BAO and $H_{0}$ to better understand their impacts on the estimation of the cosmological parameters.

\subsection{CMB data}

We use the CMB data from the latest and final Planck data release \citep{Planck_2018_6}. Its contribution in the likelihood analysis is expressed in terms of the compressed form with CMB shift parameters \citep{Wang_2007, Wang_2013}:
\begin{eqnarray}
    R &\equiv&\sqrt{\Omega_{m}H_{0}^{2}}r(z_{*})/c, \\
    l_{a} &\equiv& \pi r(z_{*})/r_{s}(z_{*}),
\end{eqnarray}
where $r_{s}(z)$ is the comoving sound horizon at redshift $z$, and $z_{*}$ is the redshift to the photon-decoupling surface. These two CMB shift parameters together with $\omega_{b}=\Omega_{b}h^{2}$ and spectral index of the primordial power spectrum $n_{s}$ can give an efficient summary of CMB data for the dark energy constraints.

The comoving sound horizon is given by
\begin{eqnarray}
    r_{s}(z)&=&\int_{0}^{t}\frac{c_{s}dt'}{a}=\frac{c}{H_{0}}\int_{z}^{\infty}dz'\frac{c_{s}}{E(z')} \nonumber \\ 
            &=& \frac{c}{H_{0}}\int_{0}^{a}\frac{da'}{\sqrt{3(1+\bar{R_{b}}a')a'^{4}E^{2}(z')}}.
\end{eqnarray}
The radiation term in the expression of $E(z)$ for the CMB data analysis shouldn't be ignored. It can be determined by the matter-radiation equality relation $\Omega_{r}=\Omega_{m}/(1+z_{\text{eq}})$, and $z_{\text{eq}}=2.5\times10^{4}\omega_{m}(T_{\text{CMB}}/2.7\text{K})^{-4}$, where $\omega_{m}=\Omega_{m}h^{2}$. The sound speed is $c_{s}=1/\sqrt{3(1+\bar{R_{b}}a)}$, with $\bar{R_{b}}a=3\rho_{b}/(4\rho_{r})$, and $\bar{R_{b}}=31500w_{b}(T_{\text{CMB}}/2.7\text{K})^{-4}$. We assume the CMB temperature $T_{\text{CMB}}=2.7255\text{K}$.

The redshift $z_{*}$ can be calculated by the fitting formula \citep{Hu_1996}:
\begin{equation}
    z_{*}=1048[1+0.00124\omega_{b}^{-0.738}][1+g_{1}\omega_{m}^{g_{2}}],
\end{equation}
where
\begin{equation}
    g_{1}=\frac{0.0783\omega_{b}^{-0.238}}{1+39.5\omega_{b}^{0.763}}, \qquad g_{2}=\frac{0.560}{1+21.1\omega_{b}^{1.81}}.
\end{equation}

The Planck data we use for the analysis include temperature and polarization data, as well as CMB lensing. In particular, we use the \sloppy{base$\_$plikHM$\_$TTTEEE$\_$lowl$\_$lowE$\_$lensing} in the \textit{base} MCMC chain, and \sloppy{base$\_$omegak$\_$plikHM$\_$TTTEEE$\_$lowl$\_$lowE$\_$lensing} in the \textit{base\_omegak} MCMC chain to get two versions of the compressed Planck CMB data for a spatailly-flat and non-flat model respectively. The final result is expressed in terms of a data vector ${\mathbf{v}}=(R, l_{a},  \omega_{b}, n_{s})^{t}$ and their covariance matrix.

For a flat universe, this data vector is
\begin{equation}
    \mathbf{v}=\begin{pmatrix}
    1.74963 \\
    301.80845 \\
    0.02237 \\
    0.96484\\
    \end{pmatrix}
\end{equation}
and the covariance matrix is 
\begin{eqnarray*}
    C_{\mathbf{v}}&=&10^{-8}\times \\
    &&\begin{pmatrix}
     1598.9554~~  17112.007~~ -36.311179~~ -1122.4683 \\
     17112.007~~  811208.45~~ -494.79813~~ -11925.120 \\
    -36.311179~~ -494.79813~~  2.1242182~~ 23.779841 \\
    -1122.4683~~ -11925.120~~  23.779841~~ 1725.4040 \\
    \end{pmatrix}.
\end{eqnarray*}
For a non-flat universe, the corresponding data vector and covariance matrix are
\begin{equation}
    \mathbf{v}=\begin{pmatrix}
    1.74451 \\
    301.76918  \\
    0.022483 \\
    0.96881 \\
    \end{pmatrix}
\end{equation}
\begin{eqnarray*}
    C_{\mathbf{v}}&=&10^{-8}\times \\
    &&\begin{pmatrix}
     2556.7782~~  23212.222~~ -57.345815~~ -1847.3003 \\
     23212.222~~  830122.02~~ -628.56261~~ -17499.134 \\
    -57.345815~~ -628.56261~~  2.5300094~~ 39.666623 \\
    -1847.3003~~ -17499.134~~  39.666623~~ 2225.2344 \\
    \end{pmatrix}.
\end{eqnarray*}
Note that the non-flat compression of the Planck CMB data has larger variance than the flat counterpart due to the added degree of freedom from $\Omega_{k}$. This compression of CMB data forms the so-called distance priors \citep{Wang_2007}. The compression of CMB data presented here is not exclusive. Examples from other MCMC chains of the Planck 2018 release can be found in the literature, e.g. \cite{Chen_2018}. For completeness, we include the spectral index $n_{s}$ in the distance prior calculation as well. In the study of the purely geometrical expansion of the universe as in this paper, this data point can be marginalized over \citep{Wang_2016}.

\subsection{BAO data}

The BAO data represent the absolute distance measurements in the Universe. From the measurements of correlation function or power spectrum of large scale structure, we can use the BAO signal to estimate the distance scales at different redshifts. In practice, the BAO data are analyzed based on a fiducial cosmology and the sound horizon at drag epoch. For instance, in an anisotropic analysis, we can measure the comoving angular diameter distance $D_{M}(z)$ and the Hubble parameter $H(z)$ \citep{Aubourg_2015}
\begin{eqnarray}
    D_{M}(z)/r_{d} &=& \alpha_{\perp}D_{M, \text{fid}}/r_{d, \text{fid}}, \\
    D_{H}(z)/r_{d} &=& \alpha_{\parallel}D_{H, \text{fid}}/r_{d, \text{fid}},
\end{eqnarray}
where $D_{H}(z)=c/H(z)$ and $r_{d}$ is the sound horizon at the drag epoch $z_{d}$, the subscript ``fid$"$ represents the quantity in the assumed fiducial cosmology. We calculate the redshift of the drag epoch as \citep{Eisenstein_1998}
\begin{equation}
    z_{d}=\frac{1291\omega_{m}^{0.251}}{1+0.659\omega_{m}^{0.828}}[1+b_{1}\omega_{b}^{b_{2}}],
\end{equation}
where
\begin{eqnarray}
    b_{1}&=&0.313\omega_{m}^{-0.419}[1+0.607\omega_{m}^{0.674}], \\
    b_{2}&=&0.238\omega_{m}^{0.223}.
\end{eqnarray}

In this paper, we use the BAO measurements at $z=0.106$ from 6dFGS \citep{Beutler_2011}, $z=0.15$ from SDSS-Main Galaxy Sample (MGS, \cite{Ross_2015}), the final DR12 BOSS measurements at redshift $z=0.38, 0.51, 0.61$ \citep{Alam_2016}, $z=1.52$ from the eBOSS QSO sample \citep{Ata_2018}, and the Ly$\alpha$ forest measurements from auto-correlation \citep{Delubac_2015} and cross correlation \citep{Font-Ribera_2014} from BOSS survey. 

\subsection{The Hubble constant}

The Hubble constant measures the current expansion rate of the Universe. The determination of its value and uncertainty has attracted significant attention for decades. However, its latest measurements from the distance ladder method and CMB observation of the early Universe reveal a controversial tension \citep{Freedman_2017}.  With the data from Gaia parallaxes, \cite{Riess_2018b} report a measurement of $H_{0}=73.52\pm1.62\text{ km s}^{-1}\text{Mpc}^{-1}$, in agreement with earlier measurement from Hubble space telescope $H_{0}=73.48\pm1.66\text{ km s}^{-1}\text{Mpc}^{-1}$ \cite{Riess_2018a}. However, these measurements are in significant tension with the latest measurement by the Planck team using their CMB data \citep{Planck_2018_6}, $H_{0}=67.27\pm0.602\text{ km s}^{-1}\text{Mpc}^{-1}$. This $3.6\sigma$ discrepancy has motivated a large number of papers that investigate it from different aspects, including the possible systematic error in the Planck data \citep{Addison_2016}, the sample variance of the local measurement of $H_{0}$ \cite{Wu_2017}, the statistical impact from the likelihood assumption \cite{Feeney_2017}, and possible implications of different cosmological models \cite{Valentino_2016, Valentino_2017, Zhai_2017}. 

In this paper, we apply the latest measurement from \cite{Riess_2018b}, and investigate its impact on the dark energy and cosmological parameter constraints. The combination with other probes including SNe and CMB can also provide hint for the study the of dark energy property, and examine possible systematic errors.
We constrain $H_0$ using SNe Ia, BAO, and Planck CMB distance priors, to compare with other measurements.

\subsection{Results}

\begin{table*}
\centering
\begin{tabular}{llll}
\hline
Data sample     &    $\Omega_{m}$    &     $\Omega_{k}$   & $h$   \\
\hline
SNe (stats)+CMB+BAO        ~&$0.3006\pm0.0059$   ~& $0.0003\pm0.0019$  ~& $0.6849\pm0.0063$ \\
SNe (stats)+CMB+BAO+H$_0$  ~& $0.2951\pm0.0055$ ~& $0.0017\pm0.0018$ ~& $0.6916\pm0.0060$      \\
SNe+CMB+BAO               ~& $0.3041\pm0.0062$  ~& $0.0001\pm0.0019$  ~& $0.6818\pm0.0065$ \\
SNe+CMB+BAO+H$_0$         ~& $0.2978\pm0.0058$  ~& $0.0017\pm0.0018$  ~& $0.6894\pm0.0062$ \\
\hline
\end{tabular}
\caption{Cosmological constraints for the non-flat $\Lambda$CDM model. The SNe data are flux-averaged, ``stats$"$ in the parenthesis refers to statistical error only for SNe data.}
\label{tab:LCDM}
\end{table*}

\begin{table*}
\centering
\begin{tabular}{lllll}
\hline
Data sample     &    $\Omega_{m}$    &     $\Omega_{k}$  & $w$  & $h$   \\
\hline
SNe (stats)+CMB+BAO        ~&$0.2982\pm0.0060$   ~& $-0.0012\pm0.0020$ ~& $-1.0494\pm0.0288$  ~& $0.6891\pm0.0069$ \\
SNe (stats)+CMB+BAO+H$_0$  ~& $0.2928\pm0.0056$ ~& $-0.0002\pm0.0020$ ~& $-1.0595\pm0.0284$ ~&$0.6962\pm0.0064$      \\
SNe+CMB+BAO               ~& $0.3035\pm0.0077$  ~& $-0.0001\pm0.0022$ ~& $-1.0075\pm0.0397$ ~& $0.6828\pm0.0086$ \\
SNe+CMB+BAO+H$_0$         ~& $0.2942\pm0.0068$  ~& $0.0004\pm0.0021$  ~& $-1.0413\pm0.0382$ ~& $0.6944\pm0.0078$ \\
\hline
\end{tabular}
\caption{Cosmological constraints for the non-flat $w$CDM model. The SNe data are flux-averaged. The value of $w=-1.0$ corresponds to the $\Lambda$CDM model.}
\label{tab:wCDM}
\end{table*}

\begin{table*}
\centering
\begin{tabular}{llllll}
\hline
Data sample     &    $\Omega_{m}$    &     $\Omega_{k}$  & $w_{0}$  & $w_{a}$ & $h$   \\
\hline
SNe (stats)+CMB+BAO        ~& $0.2968\pm0.0062$   ~& $0.0017\pm0.0036$ ~& $-1.1315\pm0.0736$ ~& $0.4444\pm0.3655$  ~& $0.6905\pm0.0071$ \\
SNe (stats)+CMB+BAO+H$_0$  ~& $0.2914\pm0.0058$  ~& $0.0033\pm0.0038$  ~& $-1.1540\pm0.0742$ ~& $0.5185\pm0.3660$  ~& $0.6979\pm0.0068$\\
SNe+CMB+BAO               ~& $0.3039\pm0.0079$  ~& $-0.0007\pm0.0034$  ~& $-0.9863\pm0.0989$ ~& $-0.1082\pm0.4815$ ~& $0.6823\pm0.0088$       \\
SNe+CMB+BAO+H$_0$         ~& $0.2940\pm0.0068$ ~&  $0.0008\pm0.0036$  ~& $-1.0542\pm0.0971$  ~& $0.0672\pm0.4853$  ~& $0.6944\pm0.0080$\\
\hline
\end{tabular}
\caption{Cosmological constraints for the non-flat $w_{0}w_{a}$CDM model. The SNe data are flux-averaged in the analysis.}
\label{tab:CPL}
\end{table*}

\begin{table*}
\centering
\begin{tabular}{llllll}
\hline
Data sample     &    $\Omega_{m}$   & $X(0.33)$  & $X(0.67)$ & $X(1.0)$ & $h$   \\
\hline
SNe (stats)+CMB+BAO    ~& $0.2936\pm0.0061$  ~& $0.9374\pm0.0299$  ~& $1.0286\pm0.0600$  ~& $0.7667\pm0.3272$ ~& $0.6940\pm0.007$  \\
SNe (stats)+CMB+BAO+H$_0$ ~& $0.2886\pm0.0057$  ~& $0.9283\pm0.0284$  ~& $1.0242\pm0.0588$ ~& $0.7150\pm0.3285$  ~& $0.7008\pm0.0067$ \\
SNe+CMB+BAO           ~& $0.3024\pm0.0079$ ~& $0.9915\pm0.0422$  ~& $1.0473\pm0.0861$ ~& $0.8372\pm0.3645$    ~& $0.6841\pm0.0090$           \\
SNe+CMB+BAO+H$_0$   ~& $0.2923\pm0.0068$    ~& $0.9554\pm0.0402$ ~& $1.0157\pm0.0801$ ~& $0.7357\pm0.3412$   ~& $0.6964\pm0.0078$   \\
\hline
\end{tabular}
\caption{Cosmological constraints for the model-independent parameterization of $X(z)$. The SNe data are flux averaged in the analysis.}
\label{tab:oXz}
\end{table*}

\begin{figure}[htbp]
\begin{center}
\includegraphics[width=9.0cm]{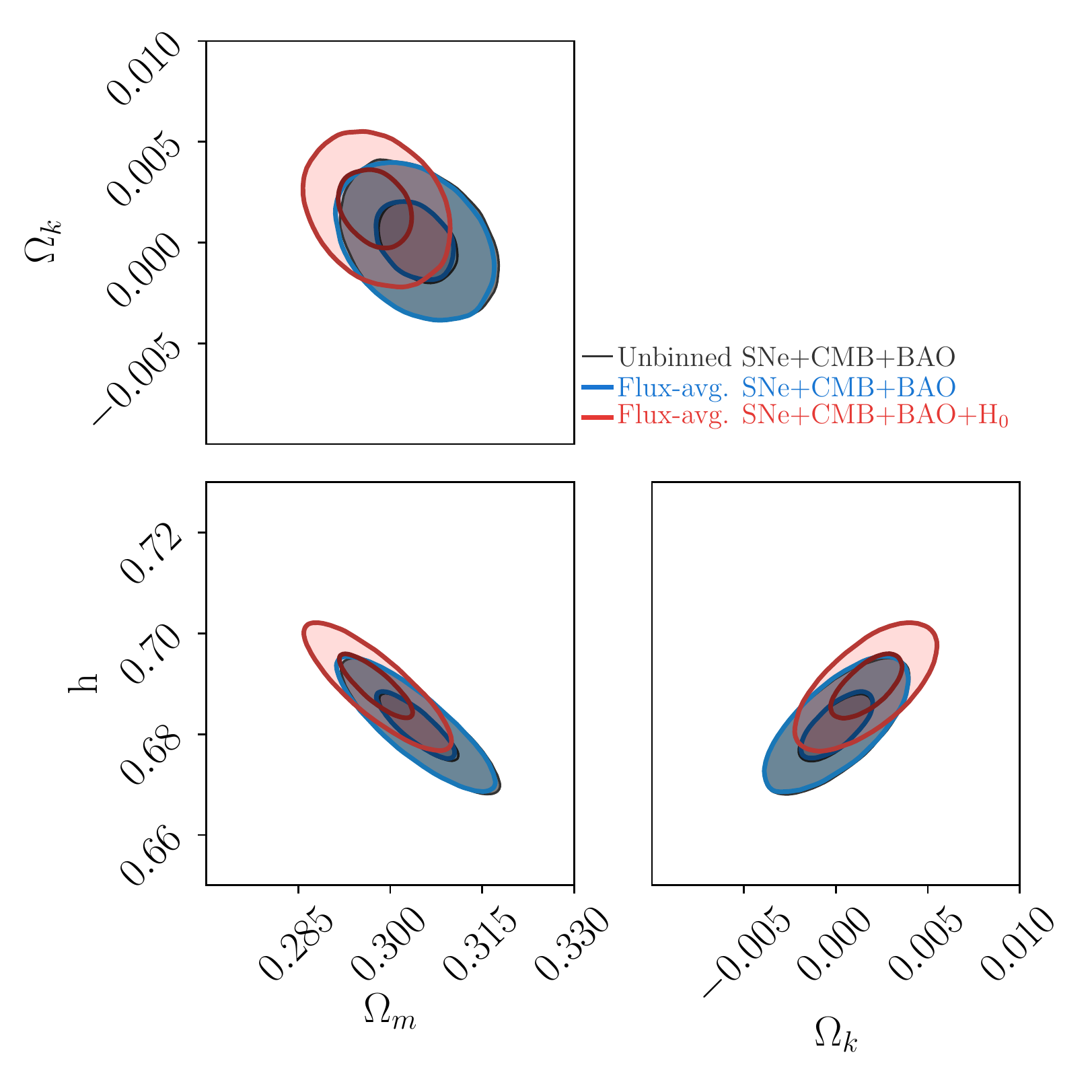}
\caption{Parameter constraints for the non-flat $\Lambda$CDM model. For comparison, the result with unbinned SNe data in combination with CMB+BAO is also shown. Due to the consistency between the flux-averaged SNe sample and the full unbinned sample, the performance of unbinned SNe+CMB+BAO is nearly identical with the flux-averaged version, so the resulting contours are indistinguishable.}
\label{fig:oLCDM_con}
\end{center}
\end{figure}

\begin{figure}[htbp]
\begin{center}
\includegraphics[width=9.0cm]{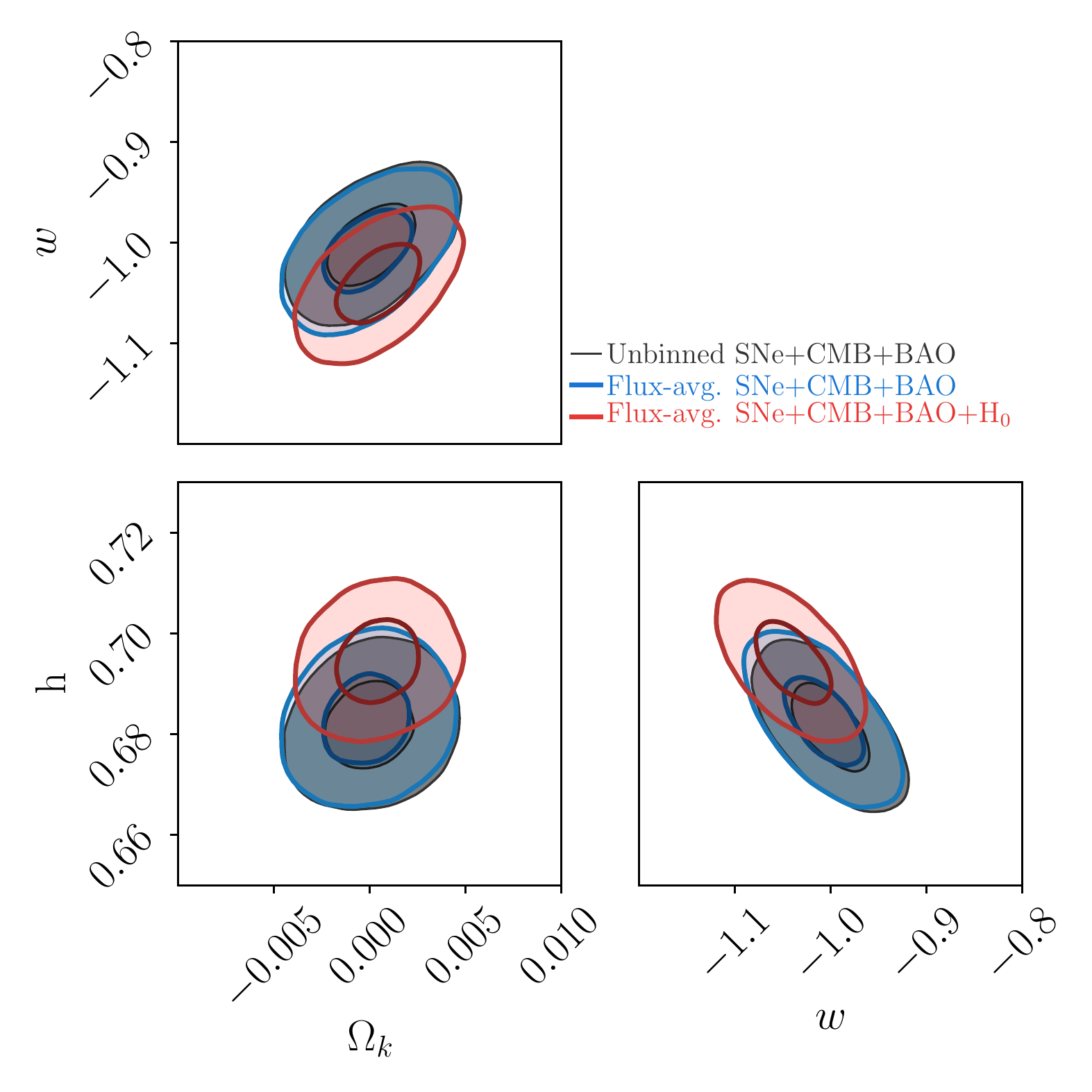}
\caption{Cosmological constraint for non-flat $w$CDM model, only parameter subset $(\Omega_{k}, w, h)$ is shown.}
\label{fig:oXCDM_con}
\end{center}
\end{figure}

\begin{figure}[htbp]
\begin{center}
\includegraphics[width=9.0cm]{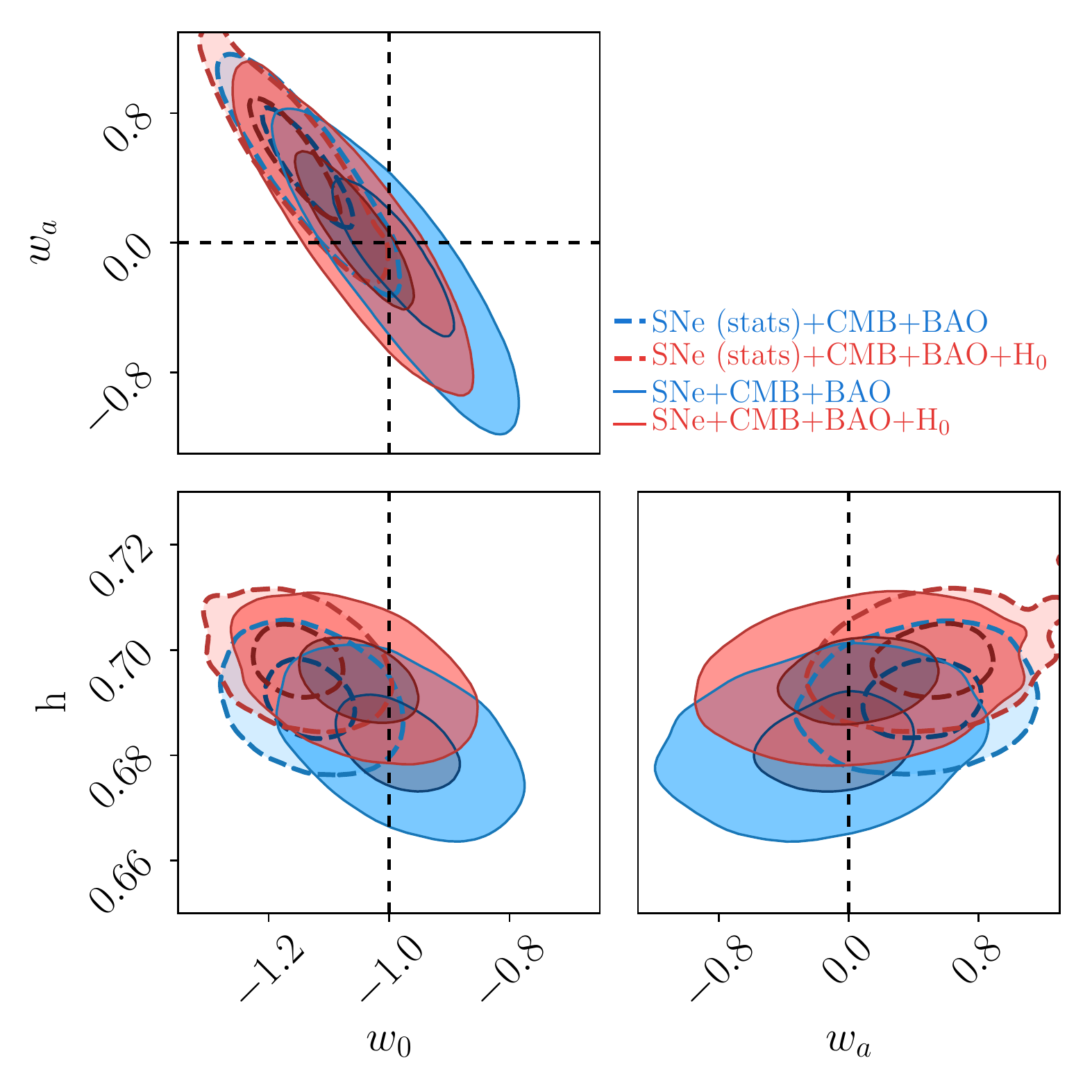}
\caption{Constraints on parameters $(w_{0}, w_{a}, h)$ for the $w_{0}w_{a}$ cosmology. SNe data are flux-averaged and the impact of systematic uncertainties is isolated. The dashed lines correspond to $w_{0}=-1.0$ and $w_{a}=0.0$. }
\label{fig:oCPL_con}
\end{center}
\end{figure}

\begin{figure*}[htbp]
\begin{center}
\includegraphics[width=18.0cm]{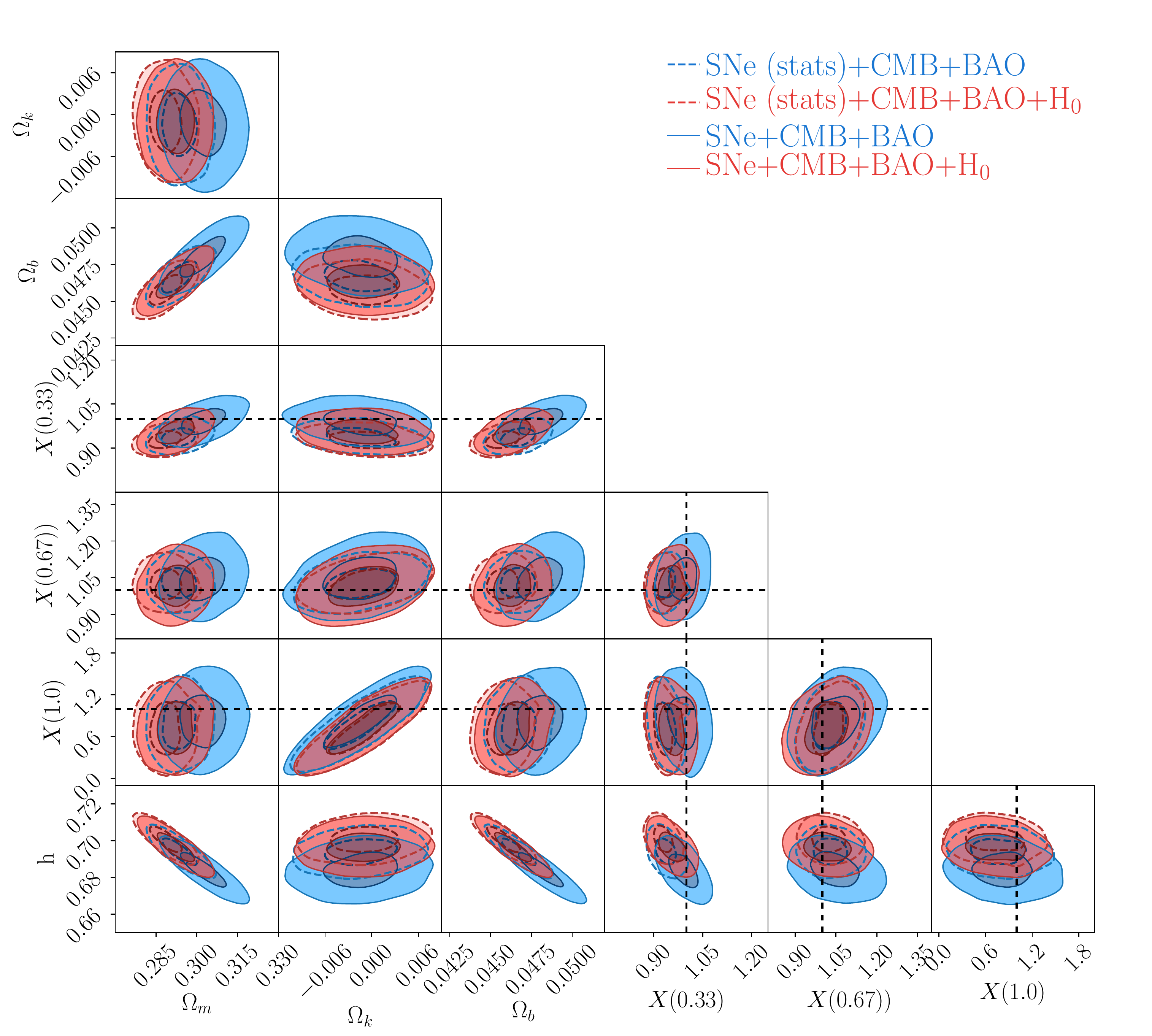}
\caption{Cosmological constrains on the model-independent parameterization of $X(z)$. The dashed lines for $X(z=0.33, 0.67, 1.0)$ correspond to values of $1.0$, i.e. cosmolgical constant. SNe data are flux-averaged in the analysis.}
\label{fig:oXz2_con}
\end{center}
\end{figure*}

\begin{figure*}[htbp]
\begin{center}
\includegraphics[width=15.0cm]{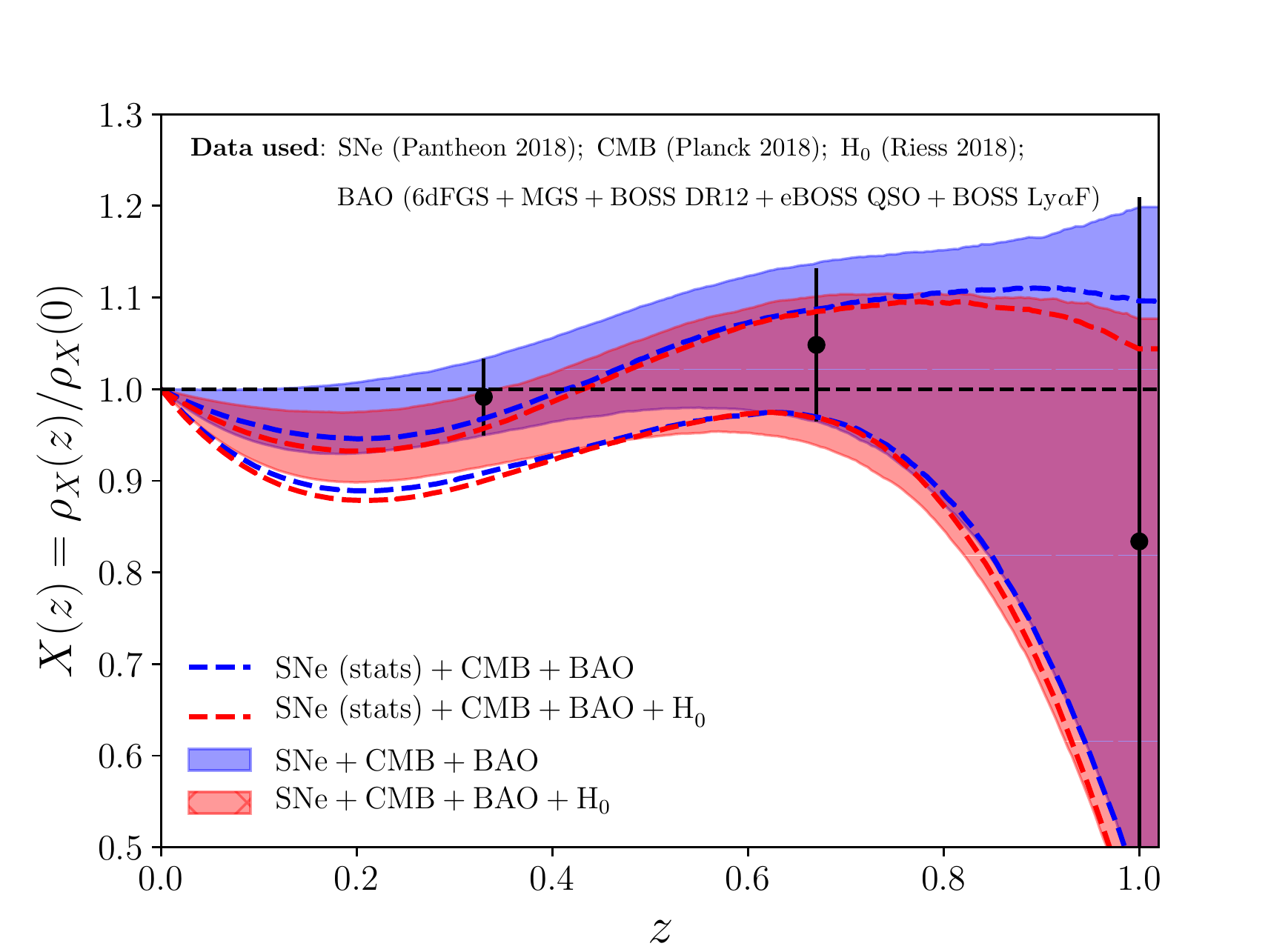}
\caption{The dark energy density function $X(z)=\rho_X(z)/\rho_X(0)$ for the model-independent parameterization. The area with shaded color or enclosed by same line style correspond to $1\sigma$ confidence level. The horizontal black dashed line correspond to $X(z)=1$, the black dots with errorbar are $1\sigma$ constraints from Flux-averaged SNe+CMB+BAO combination.}
\label{fig:oXz2_Xz}
\end{center}
\end{figure*}

Figure \ref{fig:oLCDM_con} and \ref{fig:oXCDM_con} present the constraint on the parameters for non-flat $\Lambda$CDM and $w$CDM model, the results are also summarized in Table \ref{tab:LCDM} and \ref{tab:wCDM}. Our results in Section \ref{sec:FA} show that the flux-averaged SN Ia data is in consistent with the full sample. We further investigate this effect by combining the SNe with other cosmological probes. When CMB and BAO data are used, we find that the constraints on the parameters are also identical and the resultant contours in the two figures are indeed indistinguishable. Thus we only report the results with flux-averaging of SNe in the remainder of this paper.

We also compare the effect of the systematic uncertainties from SNe on the cosmological constraints. The results in Table \ref{tab:LCDM} and \ref{tab:wCDM} show that ignoring the systematic errors in the SNe can degrade the cosmological constraints and shift the best-fit values. But the amount of change depends on cosmological models, i.e. flexibility of the model under consideration. For the simplest $\Lambda$CDM model, the constraints on $\Omega_{m}$ and $h$ can degrade by $5\%$ when the systematic errors of SNe is taken into account. However, the change in the parameters of $w$CDM is larger. The typical change of the parameter uncertainties is higher than $20\%$. The largest change is the equation of state parameter $w$ which can be affected by the SNe systematic error with an amount of $35\sim40\%$. It is due to the fact that SNe probe the late time expansion of the universe which is dominated by dark energy. Therefore it approves the necessity of the correct modeling of the systematic errors in the supernovae observations. On the other hand, the constraint on the spatial curvature $\Omega_{k}$ is barely affected by the systematic error of SNe, since most of the contribution is from CMB. 

Figure \ref{fig:oLCDM_con} and \ref{fig:oXCDM_con} also present the effect of $H_{0}$ measurement on the cosmological constraints. For instance, in the $\Lambda$CDM model, $H_{0}=68.18\pm0.65\text{ km s}^{-1}\text{Mpc}^{-1}$ from the SNe+CMB+BAO constraints, implying a $\sim3\sigma$ tension with the local measurement of $H_{0}$ from the distance ladder method. Combining all the probes with $H_{0}$ measurement, the result is shown as the pink contours in the figures. It is obvious that the addition of $H_{0}$ can shift the constraints towards the related degenerate direction. And the overall shift of the constraints is at $1\sigma$ level. For the equation of state of dark energy, adding the $H_{0}$ measurement can push the constraint to the more negative direction, implying a phantom behavior of dark energy \cite{Caldwell_2002, Caldwell_2003, Ryan_2018}. Thus understanding this $H_{0}$ tension is important in the exploration of dark energy properties.

Figure \ref{fig:oCPL_con} and Table \ref{tab:CPL} shows the constraint on the $w_{0}w_{a}$ cosmology. The effects from SNe systematic error and $H_{0}$ are both isolated in the figure. Due to the flexibility of this model, the difference between data combinations is obvious and also enhanced compared with previous simpler models. For the data combination SNe+CMB+BAO, the resulting constraints on $w_{0}$ and $w_{a}$ are consistent with a cosmological constant model. However, when the $H_{0}$ measurement is added, the current equation of state of dark energy $w_{0}$ favors a phantom value, same as the $w$CDM model. Moreover, it also slightly increases $w_{a}$, which changes the evolution of $w(z)$ in the past. The constraints without SNe systematic uncertainties are shown as the dashed contours. The results show that ignoring this systematics have similar impact on the constraint of $w_{0}$ as $H_{0}$ has, thus implying a ``degeneracy$"$ between the $H_{0}$ measurement and SNe systematic errors. The constraint on $w_{a}$ is also affected by the SNe systematic error and the change of constraint is larger than that induced by $H_{0}$. Since the systematic errors of SNe have many components, for instance the Pantheon sample has 85 separate systematic uncertainties \cite{Scolnic_2018}, understanding the correlation between the systemtaic error and the effect of $H_{0}$ on the cosmological parameter constraint is non-trivial. But it will be useful for the study of dark energy.

We present the constraints on the model-independent parametrization of $X(z)$ in Figure \ref{fig:oXz2_con} and Table \ref{tab:oXz}. This model is flexible to model the late evolution of the universe, and we show the constraints for the full parameter set in the figure. Compared with the $w_{0}w_{a}$ cosmology, the impacts from SNe systematic error and $H_{0}$ measurements are similar. In Figure \ref{fig:oXz2_Xz}, we present the evolution of the dark energy density function $X(z)=\rho_{X}(z)/\rho_{X}(0)$. The black dashed-line corresponds to cosmological constant, which is well within $1\sigma$ constraint from SNe+CMB+BAO dataset, implying that $\Lambda$CDM model is able to describe the current observations. The result also shows the effect from $H_{0}$ measurement which indicates a $1-2\sigma$ deviation from cosmological constant in the redshift range $0<z<0.33$, and this data point also forces the evolution of $X(z)$ more dynamical. This tension-motivated result is also previously investigated in \cite{Zhao_2017} with more details.

\section{Discussion and conclusion}

We have performed a systematic analysis of the latest SNe Ia sample, namely the ``Pantheon$"$ sample. We apply the flux-averaging method to this dataset to detect and minimize unknown systematic effects, and compare the cosmological constraints with the original unbinned data. The results show that the original SNe data give constraints on non-flat $\Lambda$CDM and flat $w$CDM cosmology that are consistent with the flux-averaged data. This indicates that the ``Pantheon$"$ sample has been cleaned and the systematic error from unknown systematic errors has been minimized, and that weak lensing effects are small for this sample. In addition, it supports the use of flux-averaged SN Ia data as an alternative to compress the SNe data for the cosmological usage. 
We also use these SNe data to measure the distance scale in Eq (\ref{eq:rp}) and compare with the best-fit $\Lambda$CDM model. The result shows that future supernova observation at redshift $z\sim0.1$ will be informative and can provide more constraining power on cosmological parameters. As another application of the ``Pantheon$"$ SNe sample, we use the method from \cite{Wang_2005} to measure the cosmic expansion history $H(z)$ from flux-averaged SN Ia data. The result presented in Fig \ref{fig:Ez} shows consistency with the simple $\Lambda$CDM cosmology, an interesting cross-check using a method insensitive to certain systematic uncertainties in the SNe dataset \cite{Wang_2005}. This also highlights the progress that can be made with the future SN Ia from LSST at $z<0.8$ and WFIRST at $z>1$.

In order to improve the cosmological constraints, we combine this SN Ia sample with the latest CMB and BAO data. In particular, we use the CMB distance priors that we have derived from Planck 2018 final data release, and the BAO measurements distributed in a wide redshift range. We use these data combinations and investigate the constraints on the evolution of dark energy. For flexible models like $w_{0}w_{a}$ and model-independent parametrization of dark energy density, we find that the deviation from a cosmological constant is not significant. The simplest $\Lambda$CDM model can explain the current observations with high significance. In addition, we explore the impacts from the systematic uncertainties of SNe and the local measurement of $H_{0}$ on the cosmological constraints. 

Using the combined SNe, BAO data, and Planck CMB distance priors, we find no deviation from a flat Universe dominated by a cosmological constant, and $H_0=68.4\pm 0.9\text{ km s}^{-1}\text{Mpc}^{-1}$, straddling the Planck team's measurement of $H_0=67.4\pm 0.5\text{ km s}^{-1}\text{Mpc}^{-1}$, and Riess et al. (2018) measurement of $H_0=73.52\pm 1.62\text{ km s}^{-1}\text{Mpc}^{-1}$.
Adding $H_0=73.52\pm 1.62\text{ km s}^{-1}\text{Mpc}^{-1}$ as a prior to the combined data set leads to the time dependence of the dark energy density at $z\sim 0.33$ at 68\% confidence level (see Figure \ref{fig:oXz2_Xz}). Not including the systematic errors on SNe Ia has a similar but larger effect on the dark energy density measurement.
These results may indicate possible correlations between the SNe Ia systematic errors and $H_{0}$ measurements. Understanding the detailed correlations can be challenging due to the complicated nature of SNe Ia observations, but also useful, since the $H_{0}$ measurement built on distance ladders involves the observation of supernovae. 

Identifying and correcting the potential systematic effects in the cosmological observations is crucial in the study of dark energy \cite{WangS_2013}. As future survey plans will accumulate large amounts of data, systematic uncertainties can be the limiting factor of cosmological analysis, and the correct modeling is important and necessary as we present in this paper. With more observations from ground or space \cite{DES_2018, Amendola_2018, Spergel_2015}, we can expect significant progress in our understanding of the universe in the coming decades.\\

\acknowledgments

ZZ thanks Jeremy Tinker for helpful discussions and suggestions, and Savvas Nesseris for his comments.
We acknowledge the use of the public softwares \rm{Matplotlib \citep{matplotlib},
NumPy \citep{numpy},
SciPy \citep{scipy}, and
Emcee \citep{Foreman-Mackey_2013}.
This work is supported in part by NASA grant 15-WFIRST15-0008, Cosmology with the High Latitude Survey WFIRST Science Investigation Team (SIT).
}

\bibliography{DE_GoF,software}

\end{document}